\documentclass[a4paper,aip,jcp,floatfix,unsortedaddress,superscriptaddress,
reprint]{revtex4-1}
\usepackage[utf8]{inputenc}
\usepackage{amsmath,bm}
\usepackage{graphicx}
\usepackage{color}
\usepackage{verbatim}

\begin{document}

\title{Sequence-dependent thermodynamics of a coarse-grained DNA
model}

\author{Petr \v{S}ulc}
\affiliation{Rudolf Peierls Centre for Theoretical Physics, 
 University of Oxford, 1 Keble Road, Oxford, OX1 3NP, United Kingdom}

\author{Flavio Romano}
\affiliation{Physical and Theoretical Chemistry Laboratory, 
 Department of Chemistry, University of Oxford, South Parks Road, 
 Oxford, OX1 3QZ, United Kingdom}

\author{Thomas E.~Ouldridge}
\affiliation{Rudolf Peierls Centre for Theoretical Physics, 
 University of Oxford, 1 Keble Road, Oxford, OX1 3NP, United Kingdom}

\author{Lorenzo Rovigatti}
\affiliation{Dipartimento di Fisica, Sapienza--Universit{\`a} di Roma, Piazzale
A. Moro 5, 00185 Roma, Italy}

\author{Jonathan~P.~K. Doye}
\affiliation{Physical and Theoretical Chemistry Laboratory, 
 Department of Chemistry, University of Oxford, South Parks Road, 
 Oxford, OX1 3QZ, United Kingdom}

\author{Ard~A. Louis}
\affiliation{Rudolf Peierls Centre for Theoretical Physics, 
 University of Oxford, 1 Keble Road, Oxford, OX1 3NP, United Kingdom}

\date{\today}

\begin{abstract}
We introduce a sequence-dependent parametrization for a coarse-grained DNA
model [T. E. Ouldridge, A. A. Louis, and J. P. K. Doye, J. Chem. Phys. 134,
085101 (2011)] originally designed to reproduce the properties of DNA
molecules with average sequences. The new parametrization introduces
sequence-dependent stacking and base-pairing interaction strengths chosen
to reproduce the melting temperatures of short duplexes. By developing a
histogram reweighting technique, we are able to fit our parameters to the
melting temperatures of thousands of sequences. To demonstrate the
flexibility of the model, we study the effects of sequence on: (a) the
heterogeneous stacking transition of single strands, (b) the tendency of a
duplex to fray at its melting point, (c) the effects of stacking strength in
the loop on the melting temperature of hairpins, (d) the force-extension
properties of single strands and (e) the structure of a kissing-loop
complex. Where possible we compare our results with experimental data and
find a good agreement. A simulation code called oxDNA, implementing
our model, is available as free software.
\end{abstract}

\maketitle

\section{Introduction}
Living organisms store genetic information in DNA, a double-stranded polymer composed 
of a sugar-phosphate backbone with four different kinds of bases (adenine A, thymine
T, cytosine C or guanine G) attached. The bases have highly
anisotropic mutual interactions that are responsible for the formation of
non-trivial structures, such as helical double strands, primarily through
hydrogen bonding and stacking interactions. To a first
approximation, base-pairing occurs between complementary base pairs (A-T and
G-C).\cite{Saenger1984}
Given the reliability and programmability of base-pair formation, DNA is an
obvious candidate for use in self-assembly. Indeed, DNA has been exploited
as a building block for the assembly of nanostructures and active devices:
successes
include DNA computation,\cite{Adleman9} motors,\cite{Bath2005,Bath2009}
hierarchical self-assembly of tiles\cite{Winfree98} and self-assembly
of strands into large structures such as DNA origamis.\cite{Rothemund06}

Many theoretical and computational approaches have been developed to study
DNA. At the most fine-grained level, quantum chemistry calculations can be
used to study the interactions between nucleotides.\cite{Sponer08, Perez04,
Hobza99, Sponer06, Svozil10, Sponer04} While they provide valuable
information about the ground state energies at a high level of detail, they
are computationally demanding and do not allow for the study of dynamical
processes involving breaking and forming of base pairs. Molecular
simulation packages such as AMBER\cite{cornell95} or CHARMM,\cite{CHARMM}
that retain an all-atom representation of the nucleic acids but use
empirical force fields to model their interactions, are extensively used
for computational studies of both DNA and RNA as well as their interactions
with proteins.\cite{Perez2012} Although faster than quantum chemistry
methods, they still are computationally very demanding and the time scales
they can currently access are of the order of the $\mu$s, while many
biologically and technologically relevant processes happen at the ms
timescale or longer. At the moment, simulations of rare
events such as the breaking of base pairs remain at the limit of what is
possible. At the next level of complexity are coarser models of
DNA~\cite{Drukker2001, Sales-Pardo2005, Kenward2009, Ouldridge2009,
Knotts2007, Sambriski2009, Linak11, Araque11, Florescu2011,
Morriss-Andrews2010, Savin2011, Dans2010, Savelyev2009, Becker2007,
Lankasbook} that
integrate out several degrees of freedom, such as replacing a group of
atoms by a single site with effective interactions. While these models
cannot describe the system at the same level of detail as atomistic
simulations, they allow one to study much larger systems and address rare
events. Finally, continuous models of DNA~\cite{Dauxois1993, Nisoli11,
monasson99} completely neglect the detailed chemical structure but allow
for analytical treatment in the thermodynamic limit, and have been used to
study macroscopic properties such as melting temperatures or properties of
DNA under stress.\cite{Nisoli11,monasson99}

DNA nanotechnology exploits processes that include strand diffusion and
breaking and forming of base pairs. Computational methods describing such
systems must be efficient enough to access the time scales at which these
processes happen. Moreover, the coarse-grained model must be properly
designed to capture the structural, thermodynamical and mechanical
properties of DNA in both the single- and double-stranded forms. Such a
coarse-graining approach was recently used to develop the nucleotide-level
model of Ouldridge {\em et al}.,
\cite{Ouldridge_tweezers_2010,Ouldridge2011,Ouldridge_thesis} that was
subsequently successfully applied to the study of DNA nanotweezers,
\cite{Ouldridge_tweezers_2010} kissing hairpins,\cite{Romano12a} DNA
walkers,\cite{Ouldridge_thesis} the nematic transition of dense solutions
of short duplexes,\cite{CDMnematicDNA} and the formation of DNA
cruciforms.\cite{Matek} The model was designed with an ``average-base''
representation that includes specificity of base-pairing but otherwise
neglects the dependence of interactions on sequence. Consequently, the
model is suited to study processes for which sequence heterogeneity is of
secondary importance.

Nevertheless, many biological processes and technological applications of
nucleic acids rely on sequence heterogeneity. It is well-known that A-T
and G-C pairs have different relative binding strength,\cite{Saenger1984}
with the latter being significantly stronger because of the presence of
three rather than two interbase hydrogen bonds. Moreover, the stacking
interactions that drive the coplanar alignment of neighboring bases are
known to show significantly different behavior depending on
sequence.\cite{Saenger1984} Furthermore, a strand of DNA possesses
directionality, e.g. the phosphates of the backbone connect to the
$3^\prime$ and $5^\prime$ carbon atoms in the sugars. Interactions within
a strand are therefore distinct when the bases are permuted: for example,
the interaction of neighboring AT bases depends on whether the A is in the
$5^\prime$ direction with respect to the T or {\em vice versa}.
Besides thermodynamic properties, it has been observed that mechanical and
structural properties such as flexibility, helical twist and even helix type
are also influenced by the sequence.\cite{Calladine1997,Olson98,Geggier10,Basham95}
 
\begin{figure}[tb]
\centering
\includegraphics[width=8.5cm]{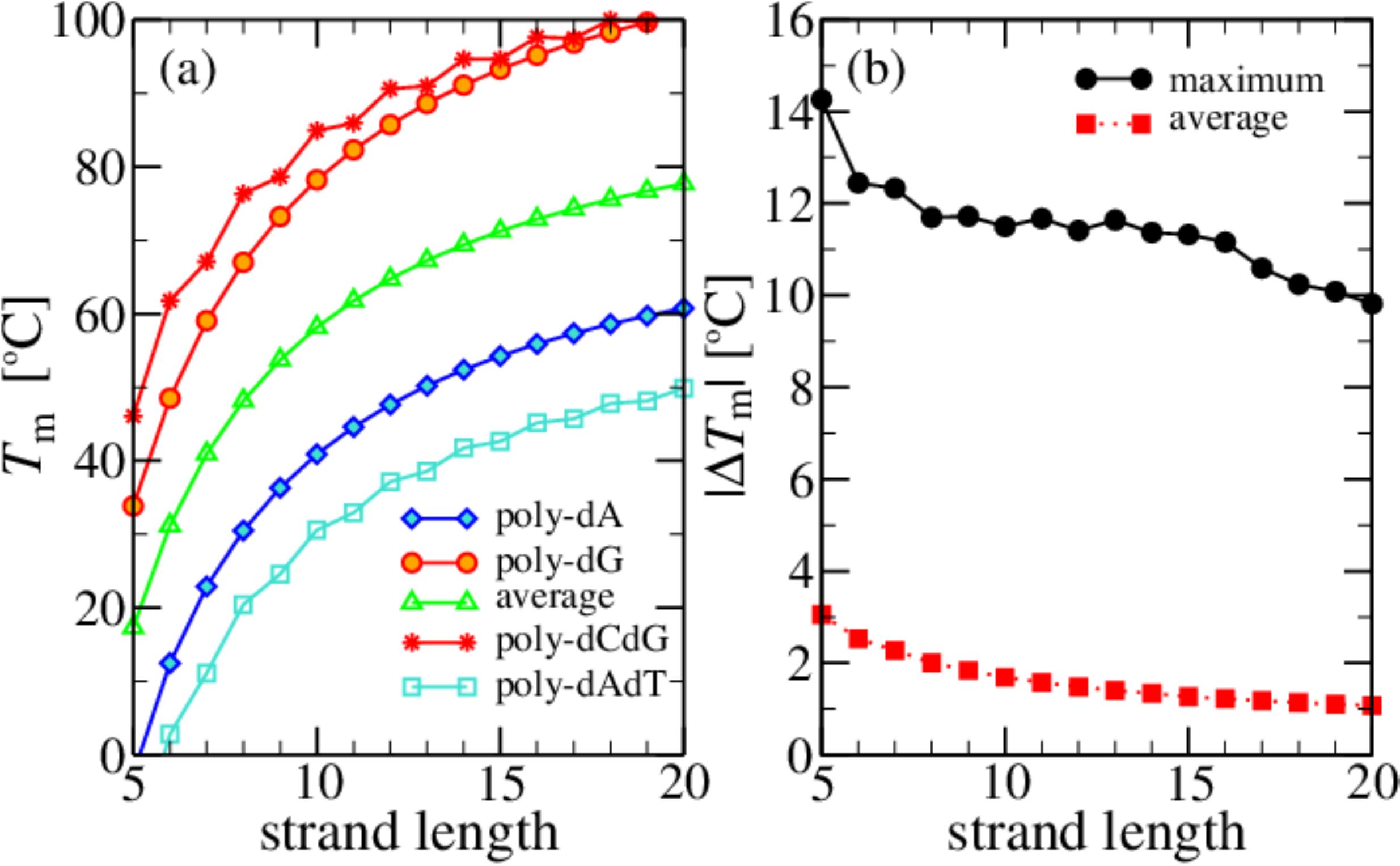}\\~\\
\caption{
(a) Melting temperatures versus duplex length as predicted by SantaLucia's
nearest neighbor model\cite{SantaLucia1998} for a duplex consisting of
poly(dA), poly(dAdT), poly(dC) or poly(dCdG) and an average sequence. (b)
Maximum (circles) and average (squares) difference in melting temperature
for strands with nucleotide positions randomly permuted.
The terminal base pairs are kept the same, thus neutralizing different end
effects. Data were generated by selecting 50000 random sequences at each
length and permuting each 5000 times. The differences show the importance
of the order of the nucleotides in the sequence.
}
\label{fig_extrem}
\end{figure}

To highlight the effects of sequence on the thermodynamics of DNA, we point
out that the melting temperature of two oligomers with the same length but
different sequences can vary by more than $50\,^{\circ}{\rm C}$, as shown
in Fig.~\ref{fig_extrem}(a) where we compare the melting temperatures of
poly(dA), poly(dG), poly(dCdG) and poly(dAdT)\footnote{We use
poly-dC,dA,dT,and dG notation for DNA sequences with repeated nucleotide
content to distinguish them from RNA sequences, which are referred to with
rC, rA, rU and rG} sequences of various lengths at an equal strand concentration of $3.36 \times 10^{-4}\,\rm{ M}$. These melting temperature
differences are only marginally diminished with increasing length and are
exploited {\em in vivo}, where, for example, it has been observed that
initiation sites of transcription are often composed of a higher than
average number of A-T pairs.\cite{cell}

Note that beside the number of A-T and G-C base pairs the actual order of
nucleotides in the sequence is also important: two sequences of the same
length and the same number of A-T and G-C base pairs can still have melting
temperatures that differ by more than $10\,^{\circ}{\rm C}$, as shown in
Fig.~\ref{fig_extrem}(b).

Given these large variations, it is important to have a model that captures
at least the thermodynamic effects of sequence. We note that some of the
other coarse-grained models of DNA that have been developed do include
sequence effects in various level of detail, including sequence dependent
base-pairing interactions\cite{Sambriski2009} and also sequence-dependent
stacking\cite{Araque11} and cross-stacking interactions.\cite{Linak11} An
extension~\cite{Ortiz11} of the model in Ref.~\onlinecite{Sambriski2009}
also has base pair deformability parametrized to the values determined by
analysis of DNA-protein crystal complexes.\cite{Olson98} In contrast to
these models, the model presented in
Refs.~\onlinecite{Ouldridge_tweezers_2010}, \onlinecite{Ouldridge2011} and
\onlinecite{Ouldridge_thesis} was specifically developed for applications
in DNA nanotechnology and was primarily designed to represent the single- to
double-stranded transition in a sufficiently physical manner. The aim of
this work is to introduce a parametrization of this model that captures the
sequence-dependence of DNA thermodynamics and use it to study sequence
effects on simple test systems.

We first present the original coarse-grained DNA model of Ouldridge {\em et
al.} \cite{Ouldridge_tweezers_2010,Ouldridge2011,Ouldridge_thesis} and then
describe the fitting procedure we developed for the sequence-dependent
interactions. We test the parametrization on melting of duplexes and
hairpins, the latter being a case to which the model was not fitted. We
then explore the flexibility of the model by studying: (a) the heterogeneous
stacking transition of single strands, (b) the tendency of a duplex to fray
at its melting point, (c) the effect of stacking strength in the loop on 
the melting temperature of hairpins, (d) the force-extension properties of
single strands and (e) the structure of a kissing-loop complex.

\section{Average base coarse-grained DNA model}
\begin{figure}[tb]
\centering
\includegraphics[width=5cm]{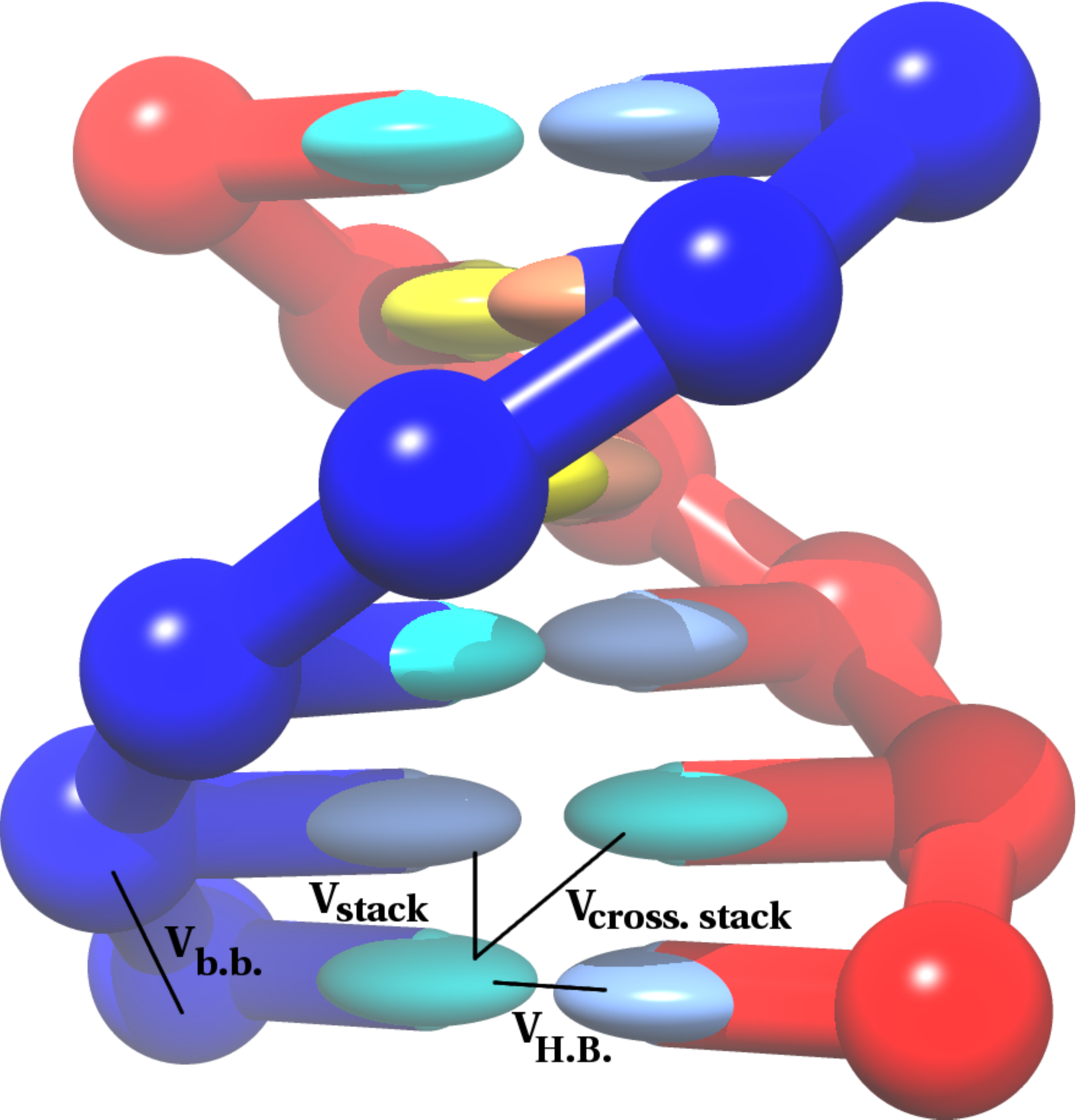}
\centering \caption{The figure shows schematically the interactions between
nucleotides in the coarse-grained DNA model for two strands in a duplex. All
nucleotides also interact with a repulsive excluded volume interactions.
The coaxial stacking interaction is not shown.
}
\label{figure_interactions}
\end{figure}

The coarse-grained DNA model, on which this work is based, is described in
detail in Refs.~\onlinecite{Ouldridge2011}
and~\onlinecite{Ouldridge_thesis}. It represents DNA as a string of
nucleotides, where each nucleotide (sugar, phosphate and base group) is a
rigid body with interaction sites for backbone, stacking and
hydrogen-bonding interactions. The potential energy of the system is
\begin{eqnarray}
 V_0 &= & \sum_{\left\langle ij \right\rangle} \left( V_{\rm{b.b.}} + V_{\rm{stack}} +
V^{'}_{\rm{exc}} \right) + \nonumber \\
    &+&  \sum_{i,j \notin {\left\langle ij \right\rangle}} \left( V_{\rm HB} +  V_{\rm{cr.st.}}  +
V_{\rm{exc}}  + V_{\rm{cx.st.}} \right) ,
 \label{eq_hamiltonian}
\end{eqnarray}
where the first sum is taken over all nucleotides that are nearest
neighbors on the same strand and the second sum comprises all remaining
pairs. The interactions between nucleotides are schematically shown in the
Fig.~\ref{figure_interactions}, and the explicit forms can be found in
Refs.~\onlinecite{Ouldridge2011} and \onlinecite{Ouldridge_thesis}. The
hydrogen bonding ($V_{\rm HB}$), cross stacking ($V_{\rm{cr.st.}}$),
coaxial stacking ($V_{\rm{cx.st.}}$) and stacking interactions
($V_{\rm{stack}}$) explicitly depend on the relative orientations of
the nucleotides as well as on the distance between interaction sites. The
backbone potential $ V_{\rm{b.b.}}$ is an isotropic spring that imposes a
finite maximum distance between neighbors, mimicking the covalent bonds
along the strand. The coaxial stacking term, not shown in the
Fig.~\ref{figure_interactions}, is designed to capture stacking
interactions between non-neighboring bases, usually on different strands.
All interaction sites also have isotropic excluded volume interactions
$V_{\rm{exc}}$ or $V^{'}_{\rm{exc}}$.

The coarse-grained DNA model of Refs.~\onlinecite{Ouldridge2011} and
\onlinecite{Ouldridge_thesis} was derived in a ``top-down'' fashion, i.e.\
by choosing a physically motivated functional form, and then focusing on
correctly reproducing the free energy differences between different states
of the system, as opposed to a ``bottom-up'' approach that starts from a more
detailed representation of DNA and typically focuses on accurate
representation of local structural details. The interactions were
originally fitted to reproduce melting temperatures of `average'
oligonucleotides, obtained by averaging over the parameters of SantaLucia's
model.\cite{SantaLucia1998} In addition, the model is fitted to reproduce
the structural and mechanical properties of double- and single-stranded DNA
such as the persistence length and the twist-modulus. The model allows for
base pairing only between Watson-Crick complementary bases, but otherwise
does not distinguish between bases in terms of interaction strengths.

The model was
fitted to reproduce DNA behavior at a salt concentration ($[\mbox{Na}^{+}] =
0.5\,$M) where the electrostatic properties are strongly screened, and it may
be reasonable to incorporate them into a short-ranged excluded volume. Such
high salt concentrations are typically used in DNA nanotechnology
applications, hence motivating this approach. It should be noted that the
model neglects several features of the DNA structure and interactions due
to the high level of coarse-graining. Specifically, the double helix in the
model is symmetrical rather than the grooves between the backbone sites
along the helix having different sizes, and all four nucleotides have the
same structure.

The main purpose of this paper is to go beyond the average sequence
parametrization by introducing sequence-dependent interaction strengths
into the model.

\section{Parametrization of sequence-dependent interactions}

We choose to perform a thermodynamic parametrization of the
sequence-dependent interactions, aiming to reproduce melting temperatures
of short DNA duplexes. We seek the parameters that best reproduce the
melting temperatures as predicted by SantaLucia's
model,\cite{SantaLucia1998} which we treat as an accurate fit to
experimental data on the melting of duplexes of different length and
sequence. We restrict sequence dependence to the strength of the base
pairing and stacking interaction terms, keeping all other parameters fixed
to the values of the original fit.

\subsection{SantaLucia's nearest-neighbor model}

In an important series of papers, SantaLucia\cite{SantaLucia1998,SantaLucia2004} summarized the results of multiple melting temperatures of DNA oligomers, and
also presented a nearest-neighbor model that reproduces the results of melting experiments
(hereafter referred to as the SL model). This popular model is the
basis of a number of widely used oligomer secondary structure and melting temperature
prediction tools.\cite{nupack,ZukerMarkham08,Markham2005,Novere01} The model assumes that DNA can exist in two
states, either single-stranded or in duplex form, and gives a standard
free-energy change of formation $\Delta G(T)$ of the duplex with respect to
the single strands as a function of temperature. The expected yields of
duplexes can then be calculated as a function of temperature
through the relation:
\begin{equation}
\frac{[AB]}{[A][B]} = \exp(-\Delta G(T) / RT),
\end{equation}
where $[A]$ and $[B]$ are molar concentrations of single strands, $[AB]$ is
the molar concentration of the duplex and $R$ is the molar gas constant.
This result assumes the system is dilute enough to behave ideally apart from associations, a
condition fulfilled in the vast majority of experiments.

The SL model assumes that $\Delta G(T)$ is a sum of contributions, one for
each base-pair step formed in a duplex with respect to the single-stranded
state, along with corrections for end effects. A base-pair step consists of
four bases; for example, the base-pair step GT/AC stands for a section of
duplex that has GT bases on one strand and AC on the complementary strand. The SL model has $10$
unique base-pair nucleotide steps: AA/TT, AT/AT, TA/TA, GC/GC, CG/CG,
GG/CC, GA/TC, AG/CT, TG/CA, GT/AC, where pairs are given in
$3^\prime$-$5^\prime$ order
along the strands.

The contribution to $\Delta G(T)$ of each term is divided into a
temperature-independent enthalpy and entropy, so that the overall form of
$\Delta G(T)$ is given by
\begin{equation}
\Delta G(T) = \Delta H - T \Delta S,
\end{equation}
with $\Delta H$ and $\Delta S$ being the (temperature-independent) sum of
the individual contributions to the enthalpy and entropy respectively. The
SL model is a {\em two-state} model, in that it considers two regions of
state space (the duplex and single-stranded states) and assumes that there
is a constant enthalpy and entropy difference between the two. In other
words, it neglects the variation in enthalpy within the bound and unbound
sub-ensembles.

The melting temperature $T_{\rm m}$ for a given
sequence is defined in the SL model as the temperature at which half of the
strands in the system are in the duplex state and the other half are in the
denatured state. Using this definition, the SL model has an average
absolute deviation of $1.6\,^{\circ}{\rm C}$ when compared to known
experimental melting temperatures of $246$ duplexes with lengths between
$4$ and $16$ base pairs.\cite{SantaLucia2004} We fit to
the $T_{\rm m}$ as predicted by the SL model, rather than having to
re-analyse the original experimental data. This choice allows us to fit to a large
ensemble of different sequences whose melting temperatures we estimate
using the SL model.

We emphasize that, in contrast to the SL model, our model itself does not
exhibit ideal two-state behavior. Although we observe a large difference in
the typical energies of single-stranded and duplex states, allowing us to
clearly differentiate the two, we also observe significant variation within
these sub-ensembles. Both single-stranded and duplex states have multiple
microscopic degrees of freedom, which respond differently to changes in
temperature. For instance, we observe fraying of duplexes
(Sec.~\ref{sec_fray}) and that the single strands undergo a stacking
transition (Sec.~\ref{sec_stacking}). The net effect is that the $\Delta
H$ and $\Delta S$ of transitions that would be inferred from our model are
not temperature independent, unlike in the SL model.

We note that other models for the prediction of DNA melting temperatures
exist, such as the recently developed nearest-neighbor model of
Ref.~\onlinecite{Ritort10}, which uses the mechanical unzipping of DNA
hairpins to infer the individual base pair step free energies. Our
parametrization procedure only requires estimates of the melting
temperature for a large set of DNA sequences and could be also used to
fit our model to the melting temperature predictions of
Ref.~\onlinecite{Ritort10}.

\subsection{Fitting of the parameters}

Our model was originally parametrized to reproduce the melting temperatures
of average sequences as predicted by the SL model.  Since the SL model is
constructed on the level of base-pair steps, it cannot be used to
differentiate between intrastrand interactions within a step: for example,
AA and TT or AG and CT.  We therefore set the stacking interaction
strengths of bases that belong to the same base-pair step to be equal in
our parametrization procedure.

To parametrize our coarse-grained DNA model's potential $V_0$
(Eq.~\ref{eq_hamiltonian}), we scale the $V_{{\rm stack}}$ and $V_{HB}$
interaction terms by the factors $\alpha_{ij}$ and $\eta_{ij}$
respectively, i.e.
\begin{eqnarray}
  V_{\rm H.B.} &\rightarrow& \alpha_{ij}V_{\rm H.B.} \label{eq_hb}
 \\
  V_{\rm stack} &\rightarrow& \eta_{ij}V_{\rm stack},
  \label{eq_hbstack}
\end{eqnarray}
where $\alpha_{ij}$ and $\eta_{ij}$ are constants for a given nucleotide
pair $ij$. There are therefore $10$ parameters $\eta_{ij}$ (as shown in
Table \ref{table_results}) and two parameters $\alpha_{\rm CG}$ and
$\alpha_{\rm AT}$ to fit. Making the cross-stacking interaction
sequence-dependent would also influence melting temperatures, but as we
will discuss later, sequence-dependent stacking and base-pairing
interactions provide enough parameters to obtain results in almost complete
agreement with the predictions given by the SL model. To fit the 12
coefficients $\eta_{ij}$ and $\alpha_{ij}$, we used a set $\mathcal{S}$ of
oligonucleotides of lengths 6, 8, 10, 12 and 18 for which we calculated the
(salt-adjusted) melting temperatures using the SL model. The set contained
2000 randomly generated sequences for each of lengths $8$, $10$, $12$, $18$
and all $4160$ sequences of length $6$. The set was then reduced to contain
only heterodimers, leaving 12\,022 sequences in total. We chose to
remove homodimers (self-complementary sequences) for convenience, because
the inference of the bulk melting temperatures from simulations of the
formation of a single duplex is different from that for heterodimers, as
discussed in Ref.~\onlinecite{Ouldridge_bulk_2010}.

We select the parameter set that
minimizes the function:
\begin{equation}
f(\alpha_{ij}, \eta_{ij}) = \sum_{s \in \mathcal{S}} \left| T_m^s(\mbox{SL}) -
T_m^s(\alpha_{ij},\eta_{ij}) \right|
\label{eq_optimize}
\end{equation}
where $T_m^s(\mbox{SL})$ is the melting temperature of the
oligonucleotide $s$ in
the set $\mathcal{S}$ as predicted by the SL model
and $T_m^s(\alpha_{ij},\eta_{ij})$ is the melting temperature predicted by our model
with sequence-dependent base pairing and stacking potentials
$\alpha_{ij}V_{\rm H.B.}$ and $\eta_{ij}V_{\rm stack}$. To accurately fit $\alpha_{ij}$ and $\eta_{ij}$, we hence need estimates of
the melting temperatures of many different sequences for many different values
of the interaction parameters.


If one simulates a system consisting of two complementary strands in the
simulation box at exactly the melting temperature then the ratio of
observed duplex states to single-stranded states
\begin{equation}
\Phi = \frac{N_{\rm duplex}}{N_{\rm single}},
\end{equation}
should be equal to $2$ for heterodimers and $1$ for homodimers. The value of $2$
for heterodimers is a correction for finite size
effects that arise when one simulates only two strands instead of a
bulk ensemble at the same average concentration.~\cite{Ouldridge_bulk_2010} The
correction assumes that the density of strands is
low enough that they behave ideally apart from association.

To calculate melting temperatures for the large set of sequences $\mathcal{S}$
we employed a histogram reweighting method.~\cite{Ferrenberg88, LandauBinderBook}
We generated once, for each duplex length considered, a set of $5000$
single-stranded and 10\,000 duplex configurations $\mathcal{C}_{\rm
single}$ and $\mathcal{C}_{\rm duplex}$. The configurations in
$\mathcal{C}_{\rm single}$ and $\mathcal{C}_{\rm duplex}$ were sampled from
the Boltzmann distribution of strands of sequence $s_0$ at the melting
temperature $T_0$ using the average parametrization (i.e., $\alpha_{ij} =
1$ and $\eta_{ij} = 1$). Simulations were performed in a cell that gave a
concentration of $3.36\times 10^{-4}\,$M for each strand. Twice as many
duplex as single-stranded states were sampled because they appear in
exactly this ratio in a simulation of two strands at the melting
temperature of a given sequence in the average model ($T_0$). Sampling was
done at sufficiently large intervals that the configurations in
$\mathcal{C}_{\rm single}$ and $\mathcal{C}_{\rm duplex}$ were
uncorrelated.

In order to find the ratio $\Phi_s(T,\alpha_{ij},\eta_{ij})$ for a sequence
$s$ at temperature $T$ with a parameter set $\alpha_{ij}$ and $\eta_{ij}$
that corresponds to a potential $V(\alpha_{ij},\eta_{ij},T)$, states in
$\mathcal{C}_{\rm single}$ and $\mathcal{C}_{\rm duplex}$ were reweighted
by the factor
\begin{equation}
 w_{l,s} \left( T,\alpha_{ij},\eta_{ij} \right) = \exp \left(\frac{V_0^{l,s_0}(T_0)}{k_{\rm B}T_0} -
 \frac{ V^{l,s}(\alpha_{ij}, \eta_{ij}, T) }{k_{\rm B}T} \right), \label{eqwls}
\end{equation}
where $V_0^{l,s_0}(T_0)$ is the energy of the $l$-th state generated at
temperature $T_0$ using the sequence $s_0$ in the average model, and
$V^{l,s}(\alpha, \eta, T)$ is the sequence-dependent potential evaluated on
the same $l$-th state for the sequence $s$. Note that both interaction
potentials are a function of temperature because the stacking interaction
term in the model is temperature
dependent.~\cite{Ouldridge2011,Ouldridge_thesis} The configurations used in
Eq.~\ref{eqwls} are generated at $T_0$ with $V_0$ and $s_0$, but each is
counted with a weight that corresponds to the desired set of new 
parameters. 

The ratio of the duplex to single-stranded states for a given temperature
$T$ and parameters $\alpha_{ij}, \eta_{ij}$ becomes 
\begin{equation} 
\Phi_s(T,\alpha_{ij},\eta_{ij}) = \frac{\sum_{l \in \mathcal{C}_{\rm duplex}} w_{l,s}
\left(T,\alpha_{ij},\eta_{ij} \right)}{\sum_{k \in \mathcal{C}_{\rm single}} w_{k,s}
\left(T,\alpha_{ij},\eta_{ij} \right)} 
\end{equation}
where the index $l$ runs through all generated duplex states while $k$ runs
through all generated single stranded states. Using this method,
$\Phi_s(T,\alpha_{ij},\eta_{ij})$ can be generated for a set of
temperatures and interpolated in order to find $T$ such that
$\Phi_s(T,\alpha_{ij},\eta_{ij}) = 2$, which is by definition the melting
temperature $T_{\rm m}$ of a given duplex.

The histogram reweighting method
assumes that the ensemble of configurations generated at temperature $T_0$
with potential $V_0$ for sequence $s_0$ is also representative of the state
space of the system at temperature $T$ and potential $V(\alpha, \eta, T)$
for sequence $s$. To check whether we included enough states, we compared the estimation of the melting temperature by
histogram reweighting of 15\,000 states to an estimation which only
used $6000$ different states. 
For a test case of 71\,000 sequences of oligonucleotide lengths $8$, $12$ and
$18$, the mean absolute deviation of the difference between the predicted
$T_{\rm m}$ was smaller than $0.1 \,^{\circ}{\rm C}$, suggesting that the
choice of 15\,000 states provides a large enough ensemble for estimating
the melting temperatures, at least on average.

%

To find a set of parameters that minimizes the function in
Eq.~\ref{eq_optimize}, we ran a simulated annealing
algorithm.\cite{LandauBinderBook} We first fitted the base-pairing
strengths $\alpha_{\rm CG}$ and $\alpha_{\rm AT}$ while holding the
stacking parameters constant. Then we fitted the 10 stacking parameters
$\eta_{ij}$ in a second step. The separate fitting of the two sets of
parameters simplifies the fitting procedure, as the converged values for
$\alpha_{ij}$ provide an initial point for the stacking parameters fitting.
It also allows us to compare the performance of a model where only the
base-pair interaction strengths are sequence-dependent to the one where both
base-pairing and stacking interactions are sequence-dependent.

We note that our fitting procedure requires the ability to efficiently
estimate melting temperatures. The histogram reweighting method, using the
generated states, takes only about 1s to calculate the melting temperature
of a given sequence. This is a huge reduction in computer time as compared
to umbrella sampling simulations,\cite{Torrie1977} which were used in the
parametrization of the original average sequence model.\cite{Ouldridge2011}
The umbrella sampling simulation samples multiple single- to
double-stranded transitions for a given oligomer and requires around two
weeks of
CPU time to calculate the melting temperature to within $0.3^{\circ}{\rm C}$
accuracy for the sequence lengths that we considered for our
parametrization. Thus our histogram re-weighting methodology provides the
crucial speed-up that made the parametrization possible.

\subsection{Parametrization results}
\label{sec_result}

While the parameters $\alpha_{\rm CG}$ and $\alpha_{\rm AT}$ were fairly
robust to details of the optimization procedure, the parameters $\eta_{ij}$
were more sensitive. In order to uniquely determine these parameters we
selected the set with the smallest average error on an additional test set
of $95\,958$ sequences that included all sequences of lengths $5$, $6$, $7$
and $8$ for which the SL model predicts a $T_{\rm m}$ greater than
$0\,^{\circ}{\rm C}$ for the concentration $3.36 \times 10^{-4}{\rm M}$,
plus a set of randomly generated sequences of lengths $10$, $12$ and $18$.
The final set of parameters $\eta_{ij}$ and $\alpha_{ij}$, as introduced in
Equations \eqref{eq_hb} and \eqref{eq_hbstack}, is shown in
Table~\ref{table_results}.

\begin{table}
\centering
\begin{tabular}{c c}
\hline
\hline
{\bf Base pairing}  &  $  \alpha_{ij} $  \\
\hline
AT & 0.8292 \\
GC & 1.1541\\
\hline
{\bf Stacking} & $ \eta_{ij}  $  \\
\hline
GC & 1.027 \\
CG & 1.059 \\
AT & 0.947\\
TA & 0.996 \\
GG, CC & 0.978 \\
GA, TC & 0.970 \\
AG, CT & 0.982 \\
TG, CA & 1.009 \\
GT, AC & 1.019 \\
AA, TT & 1.042 \\
\hline
\hline
\end{tabular}
\caption{Summary of the final parameters that were fitted to reproduce
melting temperatures of randomly chosen oligonucleotides as predicted by
the SL model. Base steps are in $3^\prime$-$5^\prime$
direction.}
\label{table_results}
\end{table}

\begin{figure}[tb]
\includegraphics[width=8.5cm]{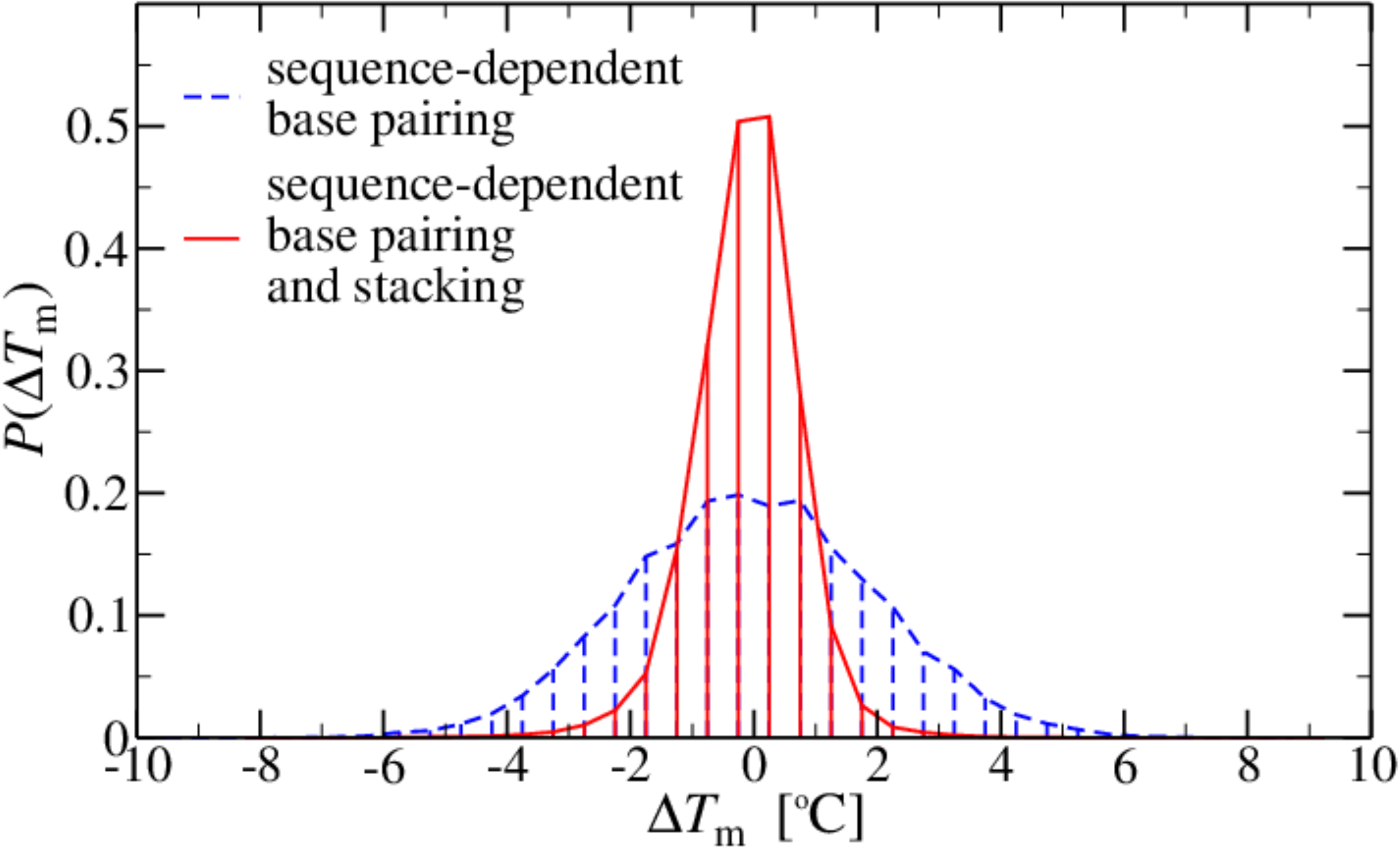}
\centering
\caption{The histogram shows the performance of the fitted DNA
coarse-grained model for the set of $95\,958$ test sequences. $\Delta
T_{\rm m}$ is the difference in the melting temperature predicted by the
coarse-grained model and by the SL model. The blue dashed curve corresponds
to a model where only hydrogen-bonding interactions were parametrized and
the red curve corresponds to the model where the stacking interactions are
also sequence-dependent (using values from Table~\ref{table_results}).}
\label{fig_performance}
\end{figure}

Figure~\ref{fig_performance} compares a histogram of the difference
\begin{equation}
 \Delta T_{\rm m} = T_m(\alpha_{ij}, \eta_{ij}) - T_{\rm m}({\rm SL}) 
\end{equation}
between the melting temperatures $T_m(\alpha_{ij}, \eta_{ij})$, calculated
by our coarse-grained model (using histogram reweighting) and the $T_{\rm
m}({\rm SL})$ of the SL model, determined for each of the $95\,958$
sequences in our test set. The dashed curve shows our model's performance
when only the base pairing interactions are sequence-dependent (parameters
$\alpha_{\rm CG}$ and $\alpha_{\rm AT}$ from Table \ref{table_results}) and
the stacking parameters $\eta_{ij}$ are all set to unity. The solid curve
shows the histogram when the melting temperatures are calculated by our
model with both hydrogen bonding and stacking sequence-dependent
parameters. The standard deviation of the distribution of $\Delta T_m$ with
sequence-dependent base-pairing and average stacking is $2 \,^{\circ}{\rm
C}$, while the standard deviation for the case where stacking is also
sequence-dependent is $0.85\,^{\circ}{\rm C}$. This compares to a standard
deviation of 8.6$\,^{\circ}{\rm C}$ for the original average-base model.
We note that although the average deviation is very small, there are a
number of melting temperatures in our set that differ significantly more
than one would expect from a Gaussian distribution with this standard
deviation. These outliers are typically highly repetitive sequences.

Since the SL model has an average absolute deviation of $1.6\,^{\circ}{\rm
C}$ when compared to experimental melting temperatures of $246$ duplexes of
lengths between $4$ and $16$, there is little point in trying to further
improve our predictions with respect to it. That it is possible to
reproduce the predictions of the SL model with our set of $12$ parameters
also implies that it would not be appropriate to introduce sequence
dependence for other terms in the interaction potential by fitting only to
$T_{\rm m}({\rm SL})$. Instead, other physical input would be needed.

It is also important to point out that, as discussed previously, by fitting
to a model which considers only base pair steps it is not possible to
distinguish between, for example, AA or TT stacking strengths, which are
known to be different.\cite{Chenw2010} Even though we treat stacking within
base pair steps equally, our method in principle allows the stacking
interaction for each individual stacked pair to be parametrized
differently. But in order to do this fitting, new experimental data is
needed. We further discuss the parametrization of stacking interactions in
section~\ref{sec_longhpin}.

\section{Tests of the parametrization}

In this Section, we test the performance of our sequence-dependent
parametrization by comparing the melting temperatures of selected duplexes,
as well as for hairpins, to which the model was not directly fitted.
 
We have also tested the structural and mechanical properties of
double-stranded DNA (away from thermodynamic transitions) on a randomly
generated sequence with around 50\% GC-content and confirmed that they are
not changed with respect to those of the original average-base
parametrization. So our double-stranded persistence length remains
approximately 125 base pairs, and the B-DNA structure produced by the model
is the same as in Refs.~\onlinecite{Ouldridge2011}
and~\onlinecite{Ouldridge_thesis}. On the other hand, the structural and
mechanical properties of single-stranded DNA properties do differ from
those of the average model, and are studied in Sec.~\ref{sec_stacking}
and~\ref{sec_pulling}.

\subsection{Duplex melting}

To further test our histogram reweighting method, we calculated several
oligomer melting temperatures using umbrella sampling Monte Carlo
simulations.\cite{Torrie1977} While histogram reweighting method estimates
the melting temperature using the same 15\,000 generated states for each
duplex length considered and extrapolates from the average-base to the
sequence-dependent potential, umbrella sampling simulations are run
separately for each sequence considered. The umbrella sampling uses the
sequence-dependent potential and is done close (within $3 \,^{\circ}{\rm
C}$) to the melting temperature of given sequence, hence providing a more
accurate estimation of the melting temperatures in our model.

The comparison between the different methods are shown in
Table~\ref{table_duplex} for a series of sequences. On average the
histogram reweighting and the umbrella sampling agree to within $0.3
\,^{\circ}{\rm C}$, which is very satisfactory. However, there is one
significant outlier, ATATAGCTATAT, for which a difference of
$2.3\,^{\circ}{\rm C}$ was obtained. One reason for the difference may be
that the melting temperature is about $16.6\,^{\circ}{\rm C}$ lower than
the melting temperature of an average strand of the same length from which
the configurations were taken for the histogram reweighting. This
difference is larger than the typical width of the melting transition
(around $10 \,^{\circ}{\rm C}$ for sequences of length 12). Moreover, the
sequence has a relatively high A-T content and may adopt structures with
significant fraying at the ends that contribute to the ensemble of
configurations for the actual strand. However, such frayed states might
have been poorly sampled when the ensemble was generated using the average
base model. For these reasons, the sampled configurations may not provide a
good representation of the true state-space of the system. Nevertheless, a
number of other sequences tested here also have melting temperatures that
differ significantly from the average sequence, without exhibiting such a
large difference in the predicted melting temperatures between the two
methods.  Although it may be true that including a significantly larger set
of states in the histogram reweighting method could reduce the errors in
these outliers, we decided not to pursue this route further, given that the
accuracy of the underlying SL model is not much different than our
parametrization errors. Should a significantly more accurate model of the
experimental data become available, however, then it may be that this point
needs to be revisited.

\begin{table}
\centering
\begin{tabular}{ccccc}
\hline
\hline
{\bf Sequence}  &   $ \mathbf{ \boldsymbol T_m(\mbox{US}) }$ &  $\mathbf{\boldsymbol T_m(\mbox{HR}) } $ & $ \mathbf{ \boldsymbol T_m( \mbox{SL} ) } $ &  $\mathbf{ \boldsymbol T_m( \mbox{SL-avg} )} $ 
\\
\hline
AAGCGT        & 38.0 & 38.2 & 39.6 & 31.2 \\
GAGATC        & 24.4 & 24.0 & 22.0 & 31.2 \\
TCTCCATG      & 44.7 & 44.6 & 44.6 & 48.2\\
CCCGCCGC      & 71.1 & 70.6 & 71.1 & 48.2 \\
ATTTATTA      & 21.2 & 21.3 & 23.9 & 48.2 \\
ATATAGCTATAT  & 47.0 & 49.3 & 48.1 & 64.7 \\
ATGCAGCTGCCG  & 74.0 & 74.3 & 72.6 & 64.7\\
GCGCAGCTGCCG  & 79.8 & 79.6 & 79.0 & 64.7\\
 \hline
 \hline
\end{tabular}
\caption{Duplex melting temperatures (shown in $\,^{\circ}{\rm C}$) as
predicted by our coarse-grained DNA model using umbrella sampling Monte
Carlo simulations (${\rm T}_{\rm m}(\mbox{US})$) and histogram reweighting
(${\rm T}_{\rm m} (\mbox{HR})$) compared to that for the SL model (${\rm
T}_{\rm m}(\mbox{SL})$). ${\rm T}_{\rm m}( \mbox{SL-avg} )$ is the melting
temperature as predicted by the averaged SL model, which depends only on
the length of the sequence. Sequences are specified in
$3^\prime$-$5^\prime$ direction.}
\label{table_duplex}
\end{table}

\subsection{Hairpin melting temperatures}
\label{hairpin_test}
We also tested our model's predictions for hairpin melting temperatures.
This provides a distinct test of the parametrized model, since the
sequence-dependent parameters were fitted to duplex melting temperatures
only. Importantly, this test also probes the quality of the model's
description of the single-stranded state, a feature often neglected in DNA
models. We test melting temperatures of 4 different hairpin-forming
sequences with different stem and loop lengths. We used strong and weak
stem sequences to highlight sequence effects.

The simulations were performed with umbrella sampling using the number of
correct base pairs in the stems as a reaction coordinate. The melting
temperature $T_{\rm m}$ is defined as the temperature at which the system
spends half of the time in the hairpin state, which is in turn defined as
the ensemble of configurations with one or more correct base pairs. In
Table~\ref{table_hairpins}, we compare our predictions for $T_{\rm m}$ with
those obtained from the SL model. The average-base parametrization was
previously found to consistently underestimate $T_{\rm m}$ for hairpins by
approximately $3\,^{\circ}{\rm C}$, but to show the correct variation with
loop and stem length.\cite{Ouldridge2011} The sequence-dependent
parametrization presented here also tends to underestimate $T_{\rm m}$ by
roughly the same amount, but the sequence effects are well captured. 

We further examine the effect of stacking on the melting temperature of
hairpins with longer loops in section \ref{sec_longhpin}, where we compare
our model with the experimentally measured influence of sequence content of
the loop on the hairpin melting temperature, an observation which is beyond
the SL model.

\begin{table}
\centering
\begin{tabular}{ccc}
\hline
\hline
{\bf Sequence} &   $ \mathbf{ \boldsymbol T_m }$ &   $ \mathbf{ \boldsymbol
T_m( \mbox{SL} ) }  $  \\
\hline
 AGCGTCACGC-$(\mbox{T})_6$-GCGTGACGCT &  86.5 &  86.7\\ 
 AGTATCAATC-$(\mbox{T})_6$-GATTGATACT &  62.2  & 64.4 \\
 AGCGTC-$(\mbox{T})_{10}$-GACGCT     &  64.5  & 67.0  \\
 AGTATC-$(\mbox{T})_{10}$-GATACT     &  44.0  & 47.3 \\
\hline
\hline
\end{tabular}
\caption{Hairpin melting temperatures (shown in $\,^{\circ}\mathrm{C}$) as
predicted by our coarse-grained DNA model ($T_{\rm m}$) compared to the
prediction by the SL model $T_m(\mbox{SL})$.
Sequences are specified in $3^\prime$-$5^\prime$ direction.}
\label{table_hairpins}
\end{table}

\section{Sequence-dependent phenomena}
\label{sec_seqdep}

To demonstrate some of the strengths and weaknesses of our new model, we
present, in this section, a series of studies of DNA systems for which
sequence plays a non-trivial role. The results were obtained from either
Monte Carlo or dynamical simulations of the model. The Monte Carlo
algorithm used is a Virtual Move Monte Carlo algorithm \cite{Whitelam2009}
and the molecular dynamics simulations were performed using a Brownian
dynamics algorithm\cite{Frenkelbook} with the thermostat as described in
Ref.~\onlinecite{Russo09}. 

\subsection{Heterogeneous stacking transition of single strands}
\label{sec_stacking}
Our model strands undergo a broad stacking transition, i.e., a transition
from a state with all or the majority of neighboring bases coplanarly
aligned to a state with disrupted alignment, as a function of
temperature.\cite{Ouldridge2011,Ouldridge_thesis} Such a transition is also
generally accepted to occur for DNA, although there is not a clear
consensus in the literature about many aspects of this
transition.\cite{Saenger1984}

\begin{figure}[tb]
\centering
\includegraphics[width=8.5cm]{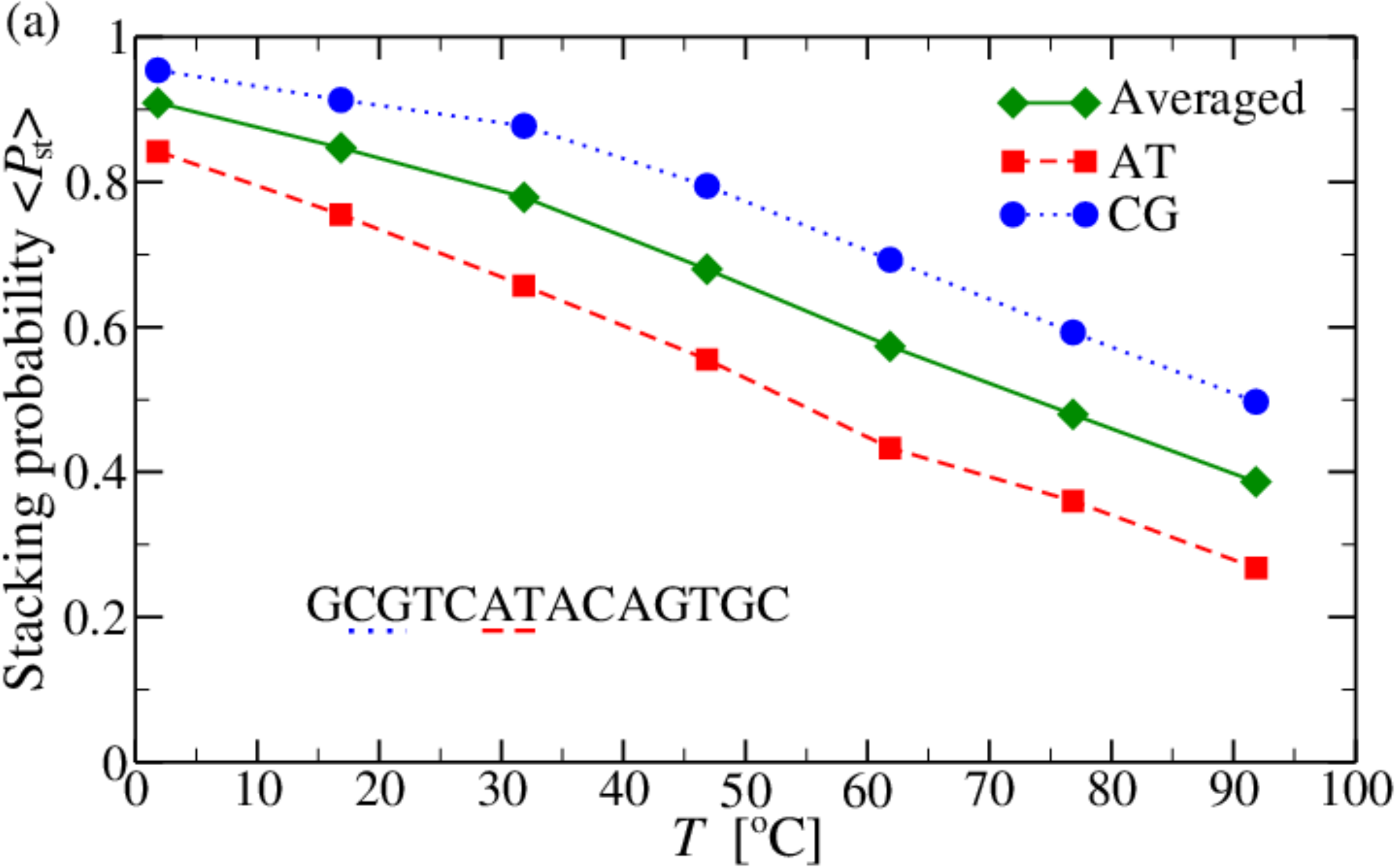}
\includegraphics[width=8.5cm]{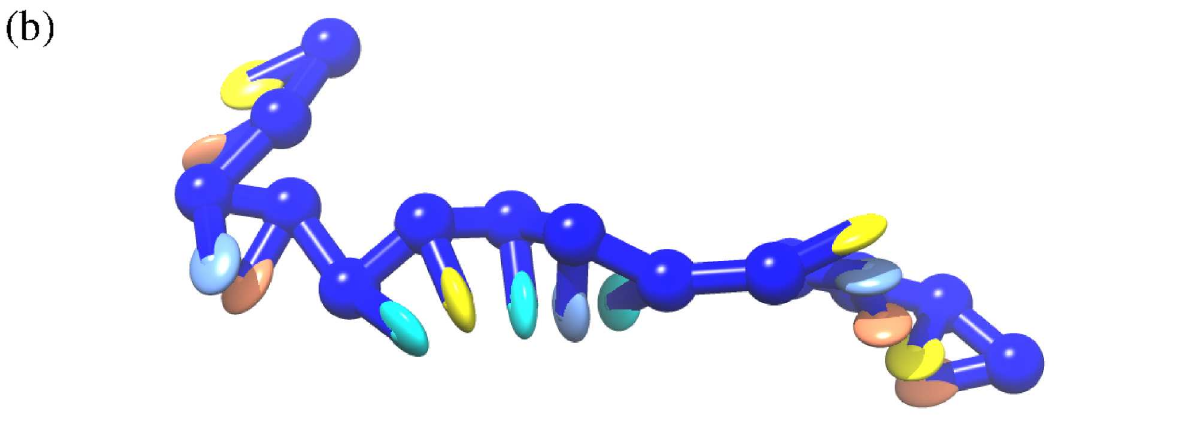}
\caption{(a) The stacking probability, calculated as the fraction of time
in the stacked state, varies with temperature and is heterogeneous along
the sequence. Circles correspond to the strongest stacking term, CG
(underscored with dotted line in sequence), while squares correspond to the
weakest stacking step, AT (underscored with a dashed line in the sequence).
Diamonds correspond to the average of all the stacking along the sequence.
(b) A typical single stranded configuration at 45$\,^{\circ}\mathrm{C}$.
The first two bases on the left are unstacked. The strand has three
stacked regions which adopt a helical geometry.
}
\label{fig_sst}
\end{figure}

To investigate the sequence dependence of stacking in our model, we ran
Brownian dynamics simulations for a 14-base single strand with sequence
GCGTCATACAGTGC (the same sequence as studied in
Ref.~\onlinecite{Holbrook1999}) at a range of temperatures. We measured the
probability that a neighbor pair stacks. Two bases are considered to be
stacked if the magnitude of their stacking interaction energy is at least
6\% of its maximal value. The choice of a cutoff is one of convenience; we
have checked that doubling it does not measurably change the results. Even
though the different stacking strengths do not vary from the average by
more than $7\%$, the effects on the stacking probabilities are still quite
significant. For example, as shown in Fig.~\ref{fig_sst}(a), the difference
between the strongest (GC) and the weakest (AT) stacking pairs is large
enough that the midpoints of the transitions are separated by about
$40\,^{\circ}\mathrm{C}$.

The structure of the single strands is also heterogeneous, consisting of
unstacked and stacked regions of various lengths, as illustrated in
Fig.~\ref{fig_sst}(b). The stacked regions adopt a helical geometry,
whereas the unstacked regions are more disordered.

The strands are also dynamically heterogeneous: over time the stacked and
unstacked regions grow and shrink, while the average probability that a
given neighboring pair of bases stack varies with temperature and position
is measured in Figure~\ref{fig_sst}(a). Mechanical and structural
properties of the single strands are therefore heterogeneous both in space
and in time.

While we are confident that the existence of significant temporal and
spatial heterogeneity in single strands is a robust qualitative prediction
of our model, given the paucity of experimental and theoretical data on the
detailed stacking interactions between individual bases, many questions
about the nature and time scales of these heterogeneities remain open.


\subsection{Hybridization free energy profiles of duplexes}
\label{sec_fray}

For our average-base parametrization, we have previously seen that duplexes
at their melting point typically have a terminal pair of bases that are
unbound. This behavior is called fraying, and it is generally thought that
the ease of fraying is sequence-dependent with A-T ends fraying more
readily.\cite{Nonin1995} To explore the fraying behavior in our model, we
study the free-energy profiles of the sequences ATATAGCTATAT, ATGCAGCTGCCG
and GCGCAGCTGCCG. Note that all three sequences have the same four central
bases but different ends.

\begin{figure}[tb]
\centering
\includegraphics[width=8.5cm]{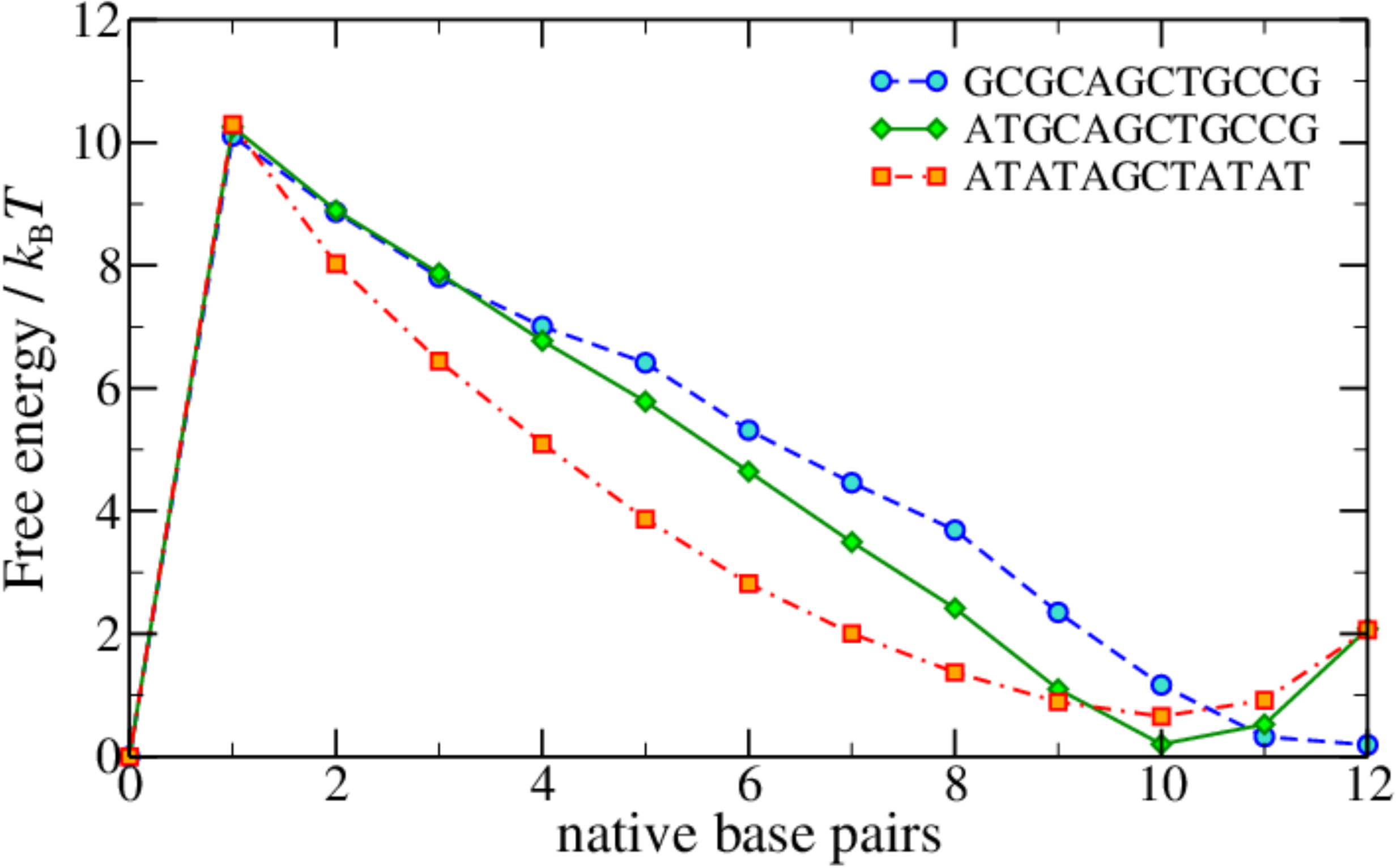}
\caption{Free energy profiles for three different duplexes of length $12$
as a function of the number of complementary (native) base pairs of the two
strands. The simulations for each duplex were run at their respective
melting temperatures, namely 48$\,^{\circ}\mathrm{C}$,
73$\,^{\circ}\mathrm{C}$ and 80$\,^{\circ}\mathrm{C}$.}
\label{fig_fray}
\end{figure}

In Fig.~\ref{fig_fray} the free energies profiles are shown as a function
of the number of the native base pairs formed between the complementary
strands. The free energies were set to be equal to 0 in the state with 0
native base pairs, i.e.\ when the duplex is melted.

Of most interest is how the most stable duplex state depends on sequence.
For the strand with two G-C ends, the free-energy minimum is a state with
all 12 bonds formed, although the free-energy cost of opening up 1
base-pair is minimal. By contrast, for the case of either one or two A-T
ends, the duplex has the lowest free energy in a state with 10 bonds
formed. Although the system pays an energetic cost for having 2 bonds
unformed, it gains entropy from this opening up of the end base pairs.
Thus, our model strands exhibit fraying, with the expected stronger
tendency to fray for duplexes with weaker A-T ends. Note that the sequence
with two A-T ends frays despite being at a significantly lower temperature
than the G-C rich sequence. Fraying has many consequences for DNA behavior.
For instance, it exposes the end bases, allowing them to take part in
reactions with other strands, which is important, for example, in a
toehold-free displacement process.\cite{Zhang2009}

Other features of note that are apparent from the free energy profiles in
Fig.~\ref{fig_fray} are the nature of the first free energy jump and the
shape of the minimum corresponding to the bound state. The fact that the
first jump is almost the same for all three sequences reflects that it is
dominated by the loss of center of mass entropy on association, which is
the same (in units of $k_{\rm B}T$) for the three systems. The shape of the
free energy minimum corresponding to the duplex highlights differences in
the ensemble of duplex states for different sequences.  For the weakest
sequence, at the melting point, the duplex can have as little as 7 base
pairs for a significant fraction of the time, and roughly with the same
probability as for it being fully closed. The most G-C rich sequence, on
the other hand, shows little tendency to fray even at its melting point and
it rarely breaks more than 3 base pairs.

\subsection{Loop sequence effect on hairpin melting temperatures}
\label{sec_longhpin}

\begin{figure}[tb]
\begin{center}
\includegraphics[width=8.5cm]{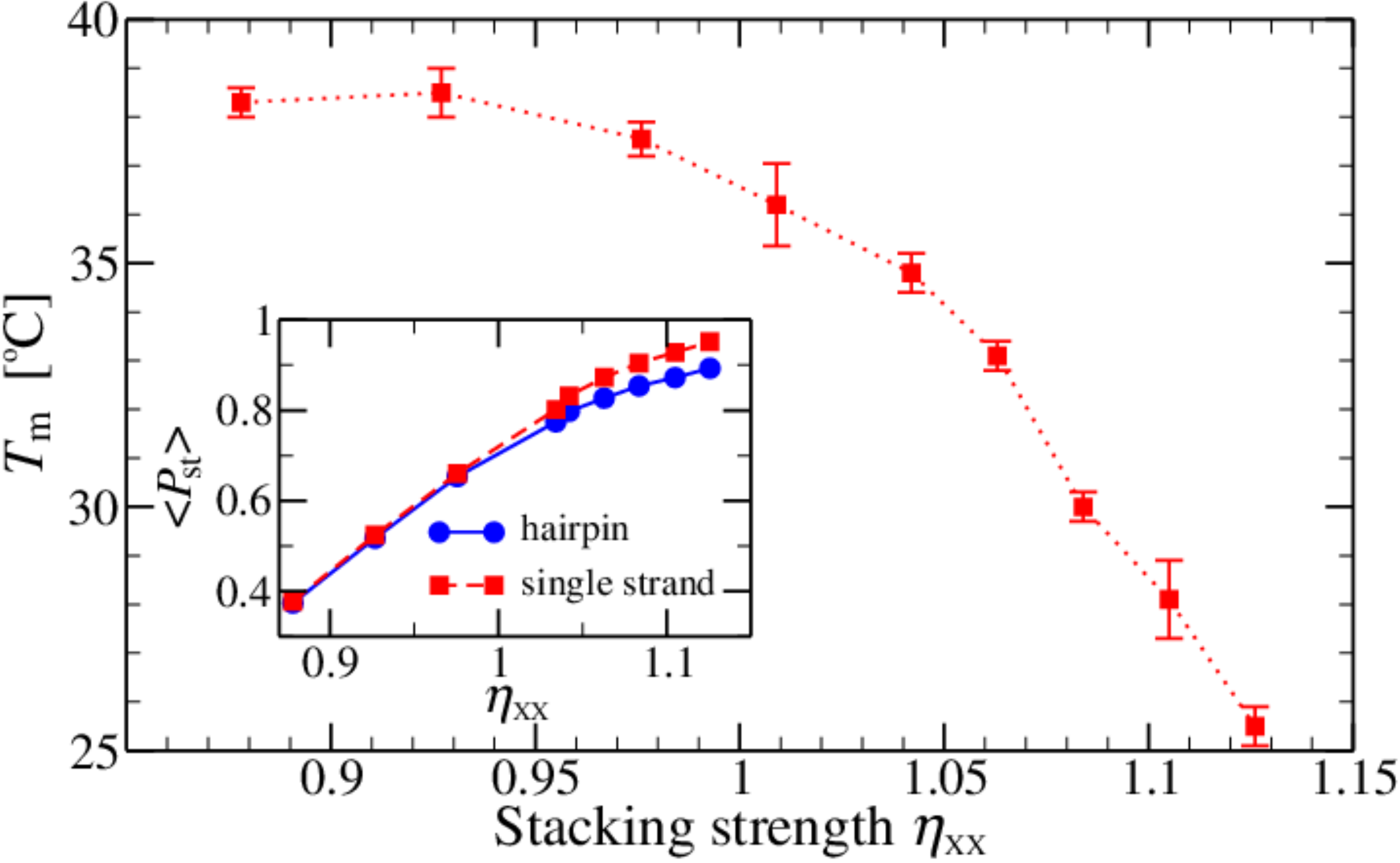}
\end{center}
\caption{
Hairpin melting temperatures as
predicted by our coarse-grained DNA model as a function of
stacking strength within the loop. We use a sequence
GGGTT-$(\mbox{X})_{25}$-AACCC, where $X$ is taken to stack as A with other
bases, and with stacking strength $\eta_{\rm XX}$ with itself.
The sequence is specified in $3^\prime$-$5^\prime$ direction. The
predicted melting temperature for the SL model is $37.8^{\circ}\mathrm{C}$.
The inset shows stacking probability $\left\langle P_{\rm st}
\right\rangle$ within the loop region in the hairpin state (circles) and
single-stranded case (squares) as a function of stacking strength
$\eta_{\rm XX}$. 
}
\label{figure_longhpins}
\end{figure}

In Section~\ref{hairpin_test}, we tested our model on melting temperatures
of hairpins with short loops of lengths 6 and 10. In the SL model, the loop
contribution to the free energy difference for closing a hairpin is
considered to be of purely entropic origin and sequence independent.
However, it was observed experimentally\cite{Goddard00} that hairpins with
the same loop lengths but different sequences have different melting
temperatures. In particular, the experiment in Ref.~\onlinecite{Goddard00}
considers sequences with the same stem sequence and loops consisting of
either poly(dA) or poly(dT). The observed difference in melting temperature
of the two different loop sequences was 4$\,^{\circ}{\rm C}$ for loop
length 12 and increased to 12$\,^{\circ}{\rm C}$ for loop length 30, with
the poly(dA) loop always having lower melting temperature. It was proposed
that the strand with a poly(dA) loop region has a higher rigidity in the
single-stranded case due to the base stacking and thus pays a larger
penalty for closing.

Although the experiments in Ref.~\onlinecite{Goddard00} were done at a salt
concentration of $0.1\,$M, lower than the $0.5\,$M to which our model was
fitted, it is instructive to see in general how stacking in the loop
influences the stability of hairpins. We calculated the melting temperature
for the sequences with the same stem sequence as in the experiment and a
range of stacking strengths in the loop. Since our model does not
distinguish between AA and TT stacking, we use an artificial base type X
that is taken to stack as A with other bases and distinctly (with stacking
strength $\eta_{\rm XX}$) with other bases of the same type X. 

The results, summarized in Fig.~\ref{figure_longhpins}, show that for
$\eta_{\rm XX} < 1$, the melting temperatures are fairly insensitive
to stacking strength whereas for $\eta_{\rm XX} \gtrsim1$, the melting
temperature starts to drop significantly with increasing stacking strength.
In the inset of Fig.~\ref{figure_longhpins} we show the average stacking
probability in the loop, compared to that of the competing single-stranded
state at the same temperature. In general, as the stacking strength
increases, the probability that a piece of single-stranded DNA has long
stacked regions also increases. The geometric constraints of the loop on
stacking therefore become more pronounced with increasing strength,
destabilizing the hairpin and leading to a drop in the melting temperature.
On the other hand, for $\eta_{xx} \lesssim1$, the stacked regions have an
average length $\langle l \rangle \lesssim 3$, which is short enough that
the hairpin geometry does not significantly affect the stacking. 


If the data of Ref.~\onlinecite{Goddard00} are to be interpreted using a
model of stacking such as ours, we would infer that poly(dA) has a very
high stacking probability at these temperatures, while poly(dT) has a
significantly lower one. But, as the inset of Fig.~\ref{figure_longhpins}
shows, we would not conclude that poly(dT) is necessarily largely unstacked. 

It is interesting to note that the stacking strength where destabilization
becomes noticeable coincides with the top end of our fitted strengths, and
that if we were to separate poly(dT) and poly(dA) stacking strengths, it
would not require an unreasonable change to give a signal of comparable
size to that reported in Ref.~\onlinecite{Goddard00}. In particular, if one
sets $\eta_{\rm AA}$ to $1.105$ and accordingly adjusts $\eta_{\rm TT}$ to
$0.979$ in order to keep the average of the two coefficients the same as
for our base-pair step parametrization, the obtained difference in melting
temperature of the hairpins with poly(dA) and poly(dT) loop is about
$9\,^{\circ}\mathrm{C}$. For these values of $\eta_{\rm TT}$ and $\eta_{\rm
AA}$ the standard deviation of melting predictions for the set of duplexes
used in testing our parametrization increases by only
$0.1\,^{\circ}\mathrm{C}$. Thus, if one wants to investigate a system where
the difference in AA and TT stacking strengths plays an important role,
these coefficients can be used. However, in the absence of a systematic
study of the effects of loop sequence on hairpin melting temperature at
high salt, we do not include differences between pairs that cannot be
distinguished by the SL model in our parametrization in
Table~\ref{table_results}.

\subsection{Force-extension curves of single strands}
\label{sec_pulling}

The mechanical properties of single strands have been experimentally
measured for both DNA and
RNA\cite{Smith1996,Mishra09,Seol2007,Seol2004,Dessinges2002,Chenw2010,Ritort10}
to characterize their average as well as base-specific properties. In
particular, qualitatively different behavior has been observed for
single-stranded poly(dT) (poly(rU) in the case of RNA) compared to poly(dA)
(poly(rC) or poly(rG) in the case of RNA); the latter exhibit significant
deviations from standard polymer models such as freely-jointed and wormlike
chains, whereas the former do not. These deviations --- concave regions
with negative curvature in the force-extension curves --- are described as
``plateaus".\cite{Mishra09,Seol2007,Chenw2010}

To investigate the effects of sequence on the mechanical properties of
single strands in our model, we simulate mechanical pulling and obtain
force-extension curves for 50-base strands at room temperature
(25$\,^{\circ}\mathrm{C}$). We consider polymers corresponding to our
weakest and strongest stacking sequences, poly(dGdA) and poly(dA), which
differ in $\eta_{ij}$ by about 7\%. We note that in
Sec.~\ref{sec_longhpin}, we used hairpin melting to distinguish AA and TT
stacking strength, but the obtained values are open to enough uncertainty
that in this section we return to our original parametrization. Our focus
here is on the qualitative effect of stacking differences, rather than
their quantitative values.

\begin{figure}[!]
\centering
\includegraphics[width=8.5cm]{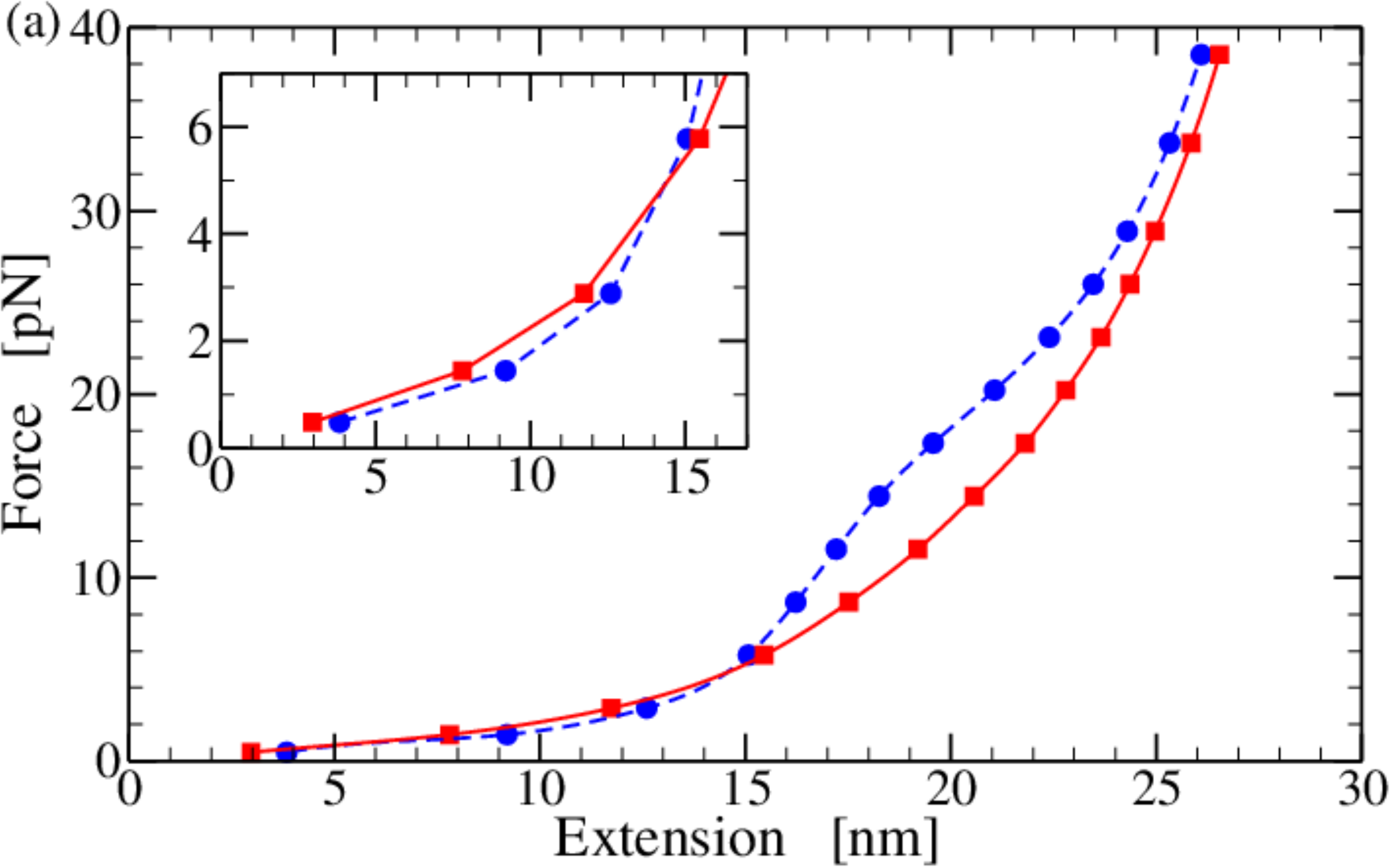}\\~\\
\includegraphics[width=8.5cm]{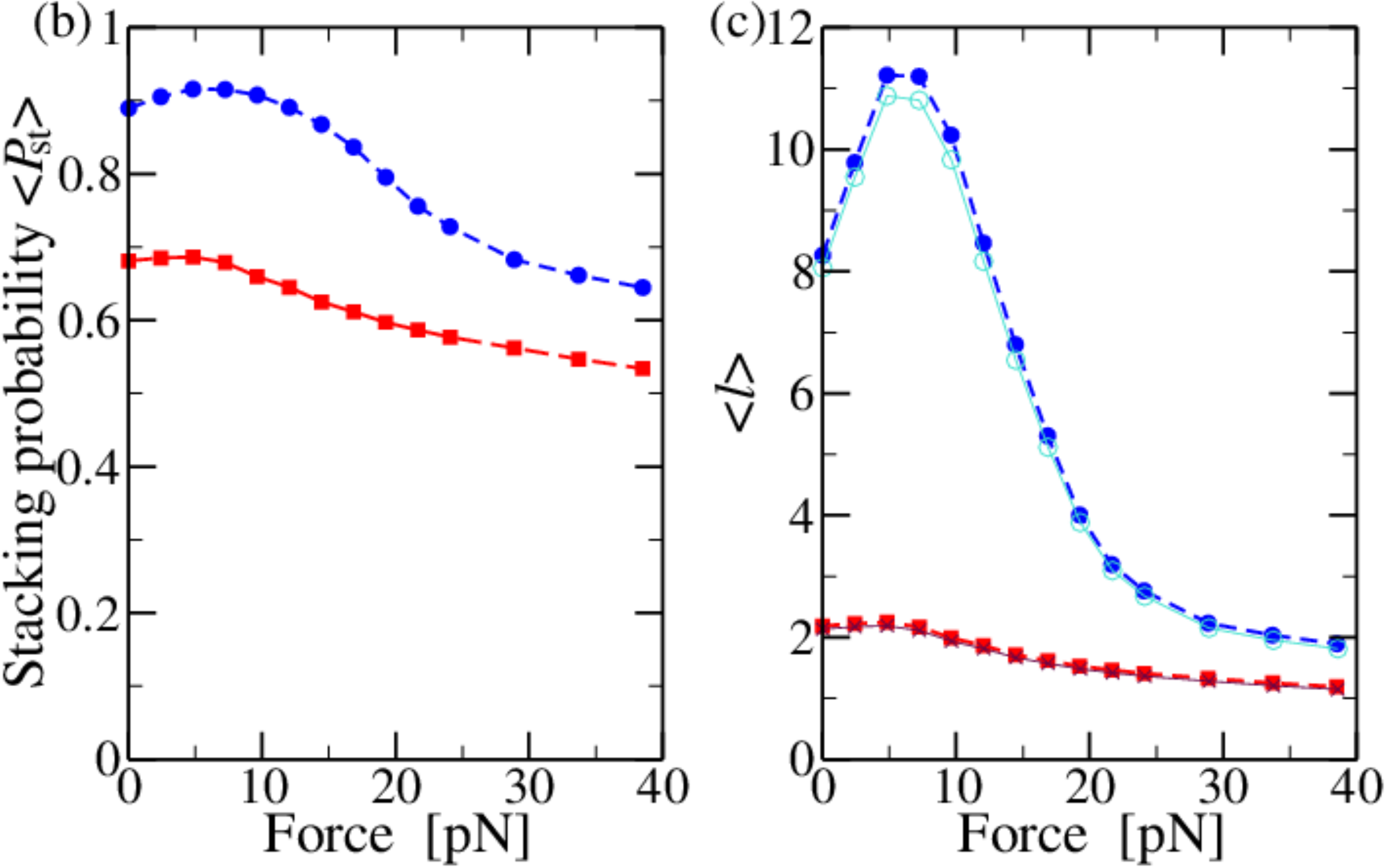}
\includegraphics[width=8.5cm]{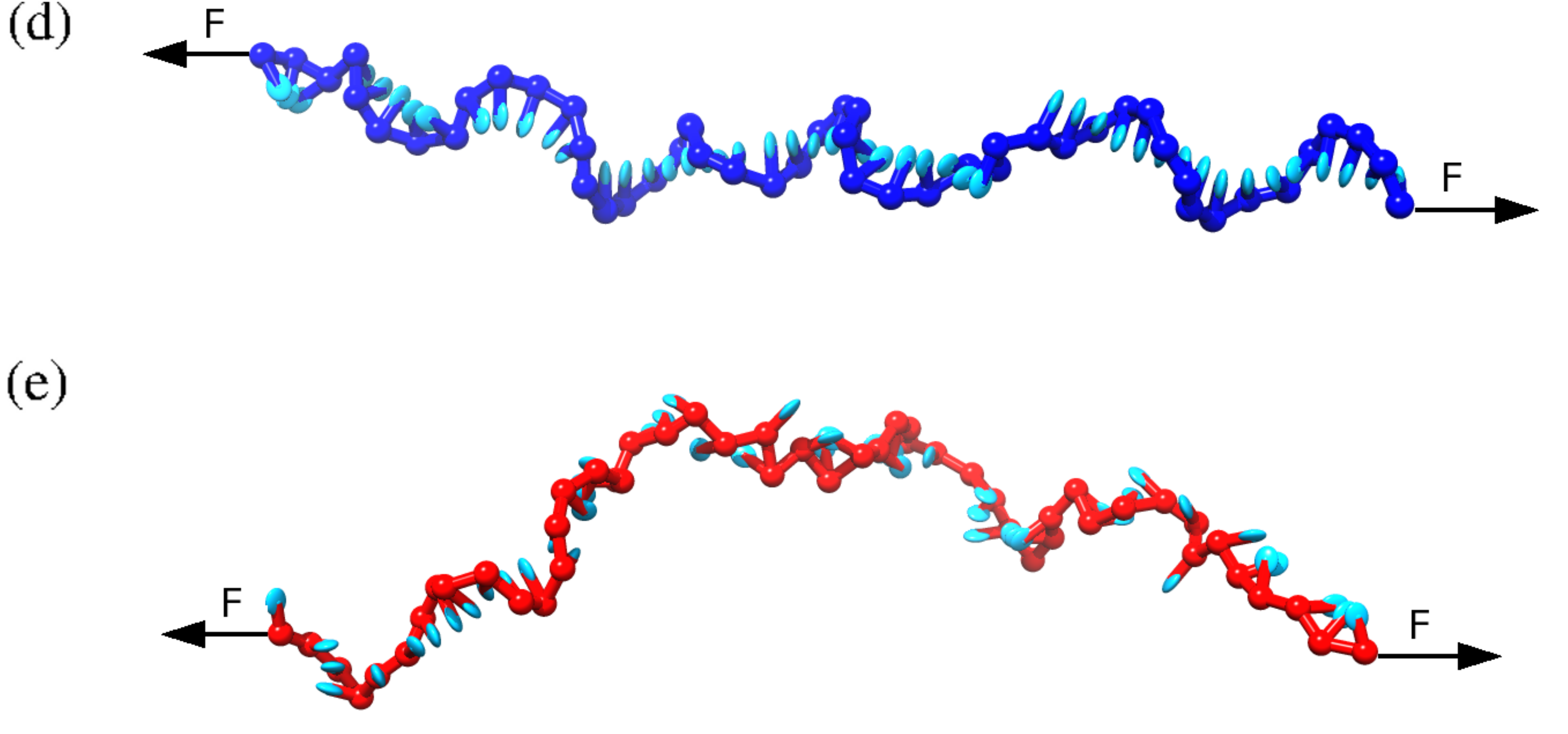}
\includegraphics[width=8.5cm]{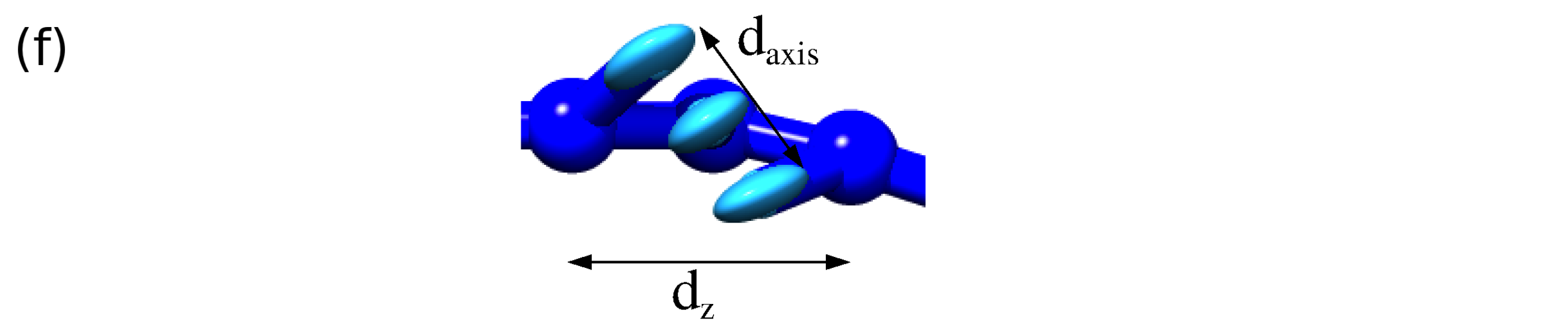}
\caption{(a) Extension of 50-nucleotide single-stranded DNA at 25$\,^{\circ}\mathrm{C}$ as a function
of applied force. In all panels, blue circles correspond to a poly(dA)
sequence (strongest stacking in our model), while red
squares correspond to a poly(dGdA) sequence (weakest stacking). The inset
in (a) shows a magnified section of the force-extension curve for low forces.
(b) Stacking probability of
a neighbor pair as a function of the applied force $F$. (c) Average length
of a stacked domain $\langle l \rangle$ as a function of applied force $F$. The open circles and crosses show $\langle l \rangle_{\rm uncoop}$
as predicted by the uncooperative stacking model (Eq.~\ref{eq_uncoop}) using $\left\langle P_{\rm st} \right\rangle$ as measured for poly(dA) and poly(dGdA) respectively. 
(d) Visualization of a 50-base long poly(dA)
ssDNA under a tension $F = 15\rm{ \,pN}$, showing multiple stacked regions with helical geometry. The arrows indicate the applied force on the 
first and the last base.
(e) Poly(dGdA) strand under a tension of 15\,pN, consisting of short stacked regions as well as unstacked ones. 
(f) Magnified section of ssDNA illustrates that three stacked bases can align with
the applied force without disrupting the stacking interaction. The contour
length ${\rm d_z}$, aligned with the force, is larger than the axial rise
${\rm d_{axis}}$.
}
\label{fig_pulling}
\end{figure}

Fig.~\ref{fig_pulling}(a) shows force-extension curves for our
strongest and weakest stacking sequences. The concave section for
strongly-stacked poly(dA) between $15$ and $25\,$pN is qualitatively
similar to the plateau-like features observed in
experiment.\cite{Seol2007,Mishra09,Chenw2010} The relatively weakly-stacked
strand, poly(dAdG), follows a convex force-extension curve which is fairly
typical of a classical homo-polymer model. 

The poly(dAdG) curve is similar to the one found for the average base
model, which in turn is in reasonable quantitative agreement with
experimental results for typical sequences. Although quantitative
comparison with experimental data for non-homopolymeric sequences, such as
$\lambda$-phage ssDNA,\cite{Smith1996} is hampered by the presence of
metastable secondary structure,\cite{Dessinges2002,Zhang2001,Montanari2000}
at tensions above about 15\,pN, where hairpins are disrupted, the extension
per base at given force in the average model is within 10\% agreement with
Ref.~\onlinecite{Smith1996}. A detailed discussion of the agreement between
the average model and experiment is given in
Ref.~\onlinecite{Ouldridge_thesis}.

To understand the difference between the two single strands in our
simulations, it is instructive to first recall that the strands consist of
dynamically changing stacked and unstacked regions, as discussed in section
\ref{sec_stacking}. When no force is applied, an unstacked region
typically has a shorter end-to-end distance than a stacked region because
it is more flexible and hence behaves more like a random coil. On the other
hand, unstacked regions also have a greater maximum extension because the
backbone is not restricted to a helical geometry as in the case of stacked
regions.

To explore the effect of pulling the structure of the single strands, we
measured the stacking probability $\left\langle P_{\rm st} \right \rangle$
and the average length $\langle l \rangle$ of contiguously stacked sections
for both strands, where a section of length $l$ consists of $l+1$ bases.
The results, as a function of applied force, are plotted in
Figs.~\ref{fig_pulling}(b) and (c). When no force is applied, the
stonger-stacking strand poly(dA) has $\langle l \rangle \cong 8$ while the
weaker-stacking strand poly(dAdG) consists mostly of short stacked regions
with average length $\langle l \rangle \cong 2$.

As shown in the inset of Fig.~\ref{fig_pulling}(a), at low forces the
stronger-stacking poly(dA) strand is more extensible than the weaker
stacking one, by as much as 20\% at $1\,$pN force. The reason for this
difference is that long stacked sections have a smaller entropic cost for
aligning with the applied force than unstacked regions do. However, as the
force increases further and the strands align more with the force, the
curves cross (at $\approx 5\,$pN), and poly(dA) becomes less extensible
because of its shorter effective contour length.

Increasing the force also leads to significant changes in the average
length of stacked regions in poly(dA). Interestingly, at low force,
the lower entropic cost for aligning of longer stacked strands leads to an
initial increase in
$\left\langle P_{\rm st} \right \rangle$ and $\langle l \rangle$ with force
(up to around $5\,$pN). 
However, as the force increases further, both $\left\langle P_{\rm st}
\right \rangle$ and $\langle l \rangle$ start to decrease because it
becomes favorable for the strand to disrupt stacking to allow for greater
extension. The reduction in stacking is particularly significant for the
poly(dA) strand over the range $15$ to $25\,$pN, the location of the
concave region in the force-extension curve. The long stacked regions are
broken down into shorter ones which facilitates an increase in the overall
length of the polymer. However, a short region of 3 bases can still align
its backbone with the force while remaining stacked, as illustrated in
Fig.~\ref{fig_pulling}(f). Therefore, even though it is progressively
reduced with force for both poly(dA) and poly(dGdA), a significant degree
of stacking is preserved even at high forces.
 
The changes in stacking hence explain the physical cause of the concave
``plateau'' region in the force-extension curve for the stronger-stacking
strand, poly(dA). It corresponds to the structural transition as the
increasing force disrupts the long stacked regions and $\langle l \rangle$
decreases. The concave segment of the force-extension curve is not present
for poly(dGdA) because the latter already consists of mostly short stacked
regions at zero force.

The differences in the structure of the poly(dA) and poly(dGdA) strands
described above are further illustrated in Figs.~\ref{fig_pulling}(d) and
(e), where snapshots of the sequences are shown for a force of $15\,$pN.
The poly(dA) strands are clearly much more stacked than the poly(dGdA)
strands are, and also more strongly aligned with the force. From this
picture one can also see why the derivative of the force-extension curve
begins to rise steeply for the poly(dA) curve around $15\,$pN: The highly
stacked strand is nearing its maximum extension, whereas the unstacked
strand is not.

It is interesting to note that a mere 20\% difference in stacking
probability between poly(dA) and poly(dGdA) at zero force causes a
significant difference in the average length of stacked regions:
$\left\langle l \right\rangle \cong 8$ versus $ \left\langle l
\right\rangle \cong 2$. This effect can be understood by considering a
simple, uncooperative model for stacking along the strand. Let $p$ be the
probability that two neighbors are stacked and $P(l)$ the probability that
a stacked cluster has length $l$.  Assuming an infinitely long polymer
chain, the probability of having a continuously stacked region of length
$l$ is $P(l)=(1-p)\,p^l$, which is the probability of having $l$ subsequent
base pairs stacked (each with probability $p$) and the $(l+1)$-th base not
stacked with the next base along the chain (which is with probability
$1-p$).
The average length $\left\langle l \right\rangle_{\rm uncoop}$ of a stacked
region in this uncooperative model can thus be obtained by summing over
$l$:
\begin{equation}
\label{eq_uncoop}
\left\langle l \right\rangle_{\rm uncoop}=
\sum_{l=0}^{\infty} l P(l) = \dfrac{p}{1-p}\:.
\end{equation}
 Since our model has low stacking cooperativity,\cite{Ouldridge2011} we can
make the approximation $p \approx \left\langle P_{\rm st}\right \rangle$.
Fig.~\ref{fig_pulling}(c) shows that this simple model compares remarkably
well with the measured values of $\left\langle l\right\rangle$.  The fact
that $\left\langle l\right\rangle$ diverges as $\left\langle P_{\rm
st}\right\rangle$ approaches 1 explains the sensitivity of the model
strands to relatively small changes in stacking propensity at large
$\left\langle P_{\rm st}\right\rangle$ and also explains the large
differences in $\left\langle l \right\rangle$ observed at zero force.


It is illuminating to compare our results to the theoretical model used by
Seol {\it et al.} in Ref.~\onlinecite{Seol2007} to explain the observed
force-extension curves of RNA. Their model makes similar physical
assumptions to the behavior of our coarse-grained model: the single strand
is split into rigid helical regions and flexible random coil regions.  Thus
the basic explanation for the plateau region is the same as in our model.
However, there are also some differences. For example, our model suggests
that absence of a plateau in the force extension curve does not necessarily
mean the absence of stacking. In fact, we have observed that short stacked
regions persist even while pulling the strand at a high force, because our
model allows for three bases to remain stacked while aligning the backbone
with the applied force, a feature that is not present in the model used in
Ref.~\onlinecite{Seol2007}.  Moreover, the concave region in the
force-extension curve interpreted with our model would indicate the
presence of a much stronger stacking propensity than the one derived in
Ref.~\onlinecite{Seol2007}. Although our description of single strands is
fairly simple, it incorporates the underlying physics of the model of
Ref.~\onlinecite{Seol2007} and in addition provides an explicit
3-dimensional representation of single-stranded nucleic acids. In summary,
We believe that the presence of concave region in the force extension curve
suggests that long stacked regions are present in the relaxed strand. This
would either indicate strong uncooperative stacking, as in our model, or
large cooperativity in stacking.

\subsection{The structure of a kissing complex}

In a recent publication,\cite{Romano12a} we investigated DNA kissing
complexes, a system where topological and geometrical frustration have
important effects, and studied the ability of the original average base
model to describe these systems. In this section, we show how the
sequence dependence of interactions can introduce non-trivial changes to
the structure of a kissing complex, with potential importance for the
operation of nanotechnological
systems.\cite{Bois2005,Dirks04,Venkataraman2007,Green2008,Yin2008,Muscat2011}

\begin{figure}[tb]
\begin{center}
\includegraphics[width=8.5cm]{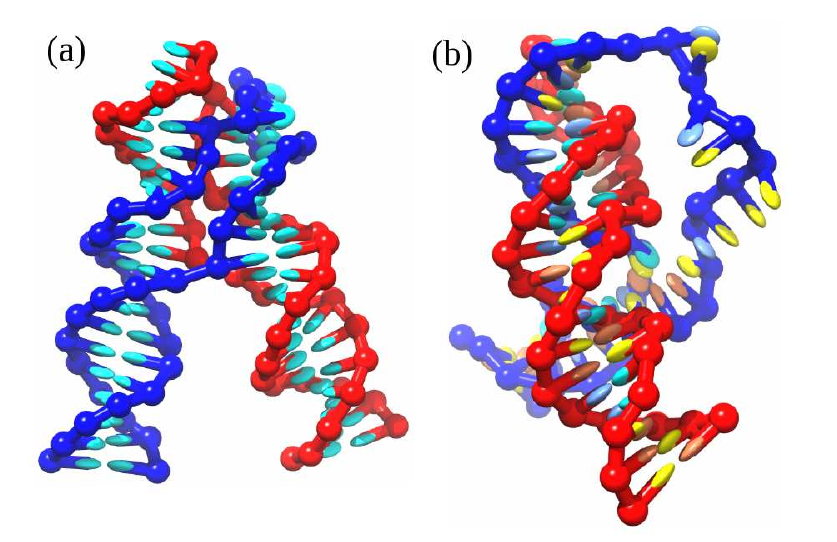}\\
\includegraphics[width=8.5cm]{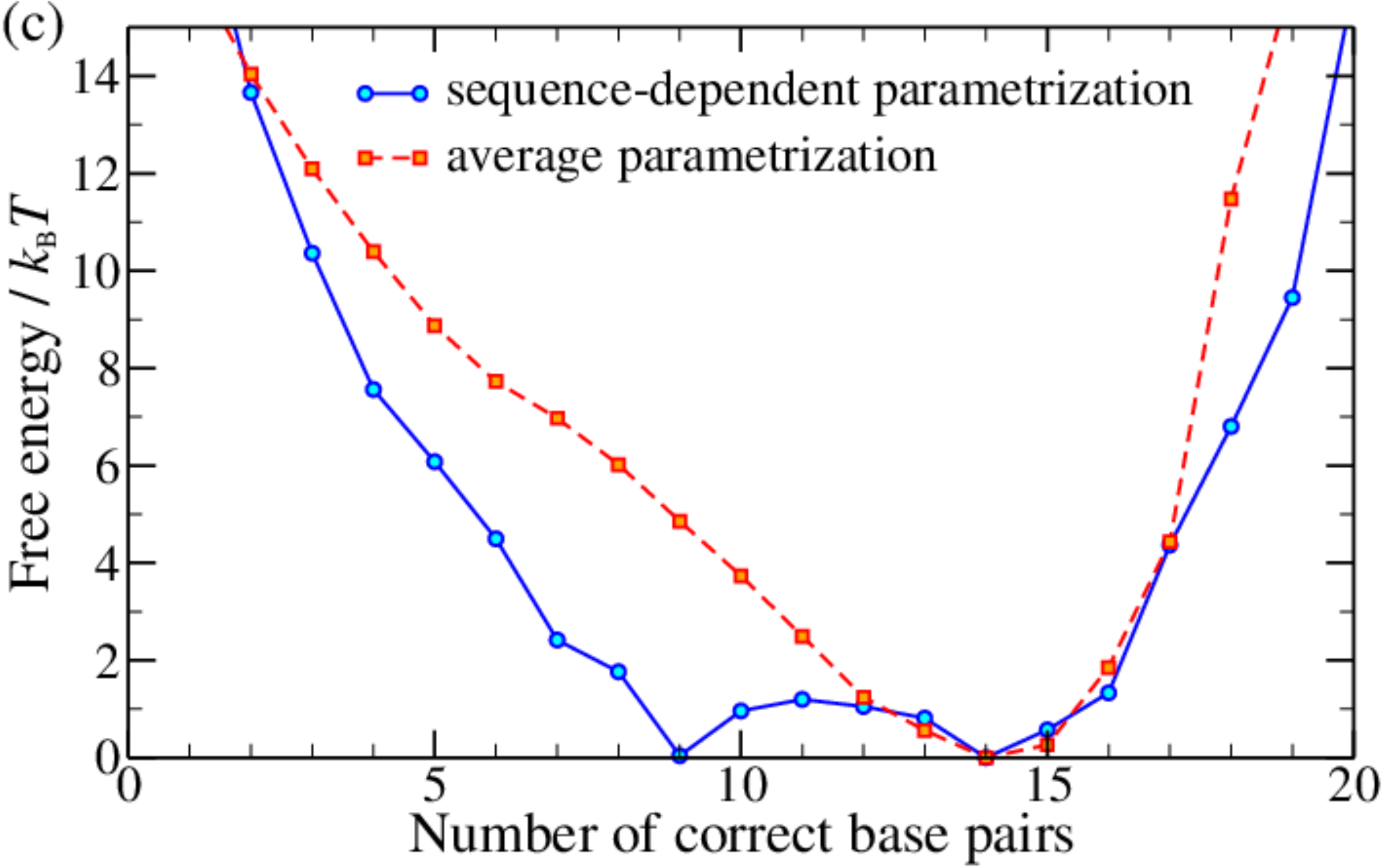}
\end{center}
\caption{
Effects of sequence dependence on the structure of kissing hairpins. (a)
Typical structure found in both the average and sequence-dependent
parametrization, with 14 intramolecular base pairs. (b) Second free energy
minimum found only in the sequence-dependent parametrization, with 9
intramolecular base pairs. Please note the exposed bases---not present in
(a)---that can be used as a toehold by the catalyst strand to initiate
displacement. (c) Free energy profile for binding with the two
parametrizations, with the sequence-dependent one exhibiting a second
minimum corresponding to the structure depicted in (b).
}
\label{fig_kissingfigure}
\end{figure}

A kissing complex is a system in which two hairpins have loop regions that
are complementary and can thus at least partially hybridize (see
Fig.~\ref{fig_kissingfigure}(a)). They are a common motif in RNA and are
expected to form in DNA
nanotechnology systems where complementary hairpins are used as fuel for
DNA nanomachines.~\cite{Bois2005,Green06a} In the experimental system
realized in Ref.~\onlinecite{Bois2005}, two strands of 40 nucleotides were
designed to be both complementary and also able to form a hairpin with a
stem of 10 base pairs. As the remaining 20-base loops are complementary to
each other, the two hairpins can form a kissing complex. The sequences
are~\cite{Bois2005}
\begin{center}
\begin{small}
\begin{verbatim}
3'-CGCAACGACG-GCTCCCCTCTTCTCATTTTA-CGTCGTTGCG-5'
\end{verbatim}
\end{small}
\end{center}
and
\begin{center}
\begin{small}
\begin{verbatim}
3'-CGCAACGACG-TAAAATGAGAAGAGGGGAGC-CGTCGTTGCG-5'
\end{verbatim}
\end{small}
\end{center}
where the hyphens separate stem and loop regions.
A dilute solution of such strands tends to form hairpins much more quickly than 
full duplexes, due to a lower kinetic
barrier for the former process. The hairpins in turn form kissing complexes, an intermediate
metastable state with respect to full hybridization that requires a
significant amount of rearrangement to transform into the full duplex. The
kinetic barrier, due to the topological frustration of the complex, is so
high that full hybridization is almost impossible. However, this barrier can be
reliably resolved by the introduction of a DNA catalyst strand, designed to
open one of the hairpins by displacement and trigger full
hybridization, thus releasing the stored free-energy.~\cite{Bois2005}

Following Ref.~\onlinecite{Bois2005}, we studied in
Ref.~\onlinecite{Romano12a} the structure of the resulting kissing
complex with the average sequence parametrization. We found that the
system typically assumed a structure with two symmetric parallel helices, as
shown in
Fig.~\ref{fig_kissingfigure}(a). However, as the loop sequences used in Ref.~\onlinecite{Bois2005} are very
asymmetric in G-C content, we expect that the average parametrization should
overestimate the
stability of the weakly bound region and conversely underestimate that of the
strongly bound, G-C-rich region. 

When we repeated the structural study with the sequence-dependent
potential, we obtained a qualitatively different result. Computing the
binding free-energy profile of the system, using the number of native base
pairs (i.e.\ base pairs that would be present in the final full duplex) as a
reaction coordinate (Fig.~\ref{fig_kissingfigure}(c)), we found a second
minimum at around nine interstrand base pairs that was not observed for the
average parametrization. A typical configuration associated with this
minimum is shown in Fig.~\ref{fig_kissingfigure}(b). It is evident that as
well as being able to form the structure with two symmetric helices, the
system is also able to adopt an alternative structure with a single
intermolecular helix that both contains the G-C-rich section and is
slightly larger than either individual helix in the two-helix form.

This competing minimum has potentially important consequences for the
nanotechnological applications of kissing hairpins.
In Ref.~\onlinecite{Bois2005}, a catalyst strand was introduced to the system in
order to facilitate full hybridization of the complex:
the strand was designed to bind to the weaker half of one of the loops, and then to
open up the hairpin by displacement.
The fact that a competing minimum exists in which the whole weaker half of the
loop is available for binding will favor this process,
as it provides a long, easily accessible toehold for displacement. Such toeholds
are known\cite{Zhang2009} to accelerate displacement reactions 
by several orders of magnitude. Our model suggests that if the strand was
instead designed to bind to the stronger half of the loop, its effectiveness
would be hindered rather than helped by the presence of the alternative minimum.
We would therefore expect such a catalyst to be less effective than the one used
in Ref.~\onlinecite{Bois2005}.

The qualitative difference between the results of the two parametrizations in
this case highlights that if one is interested in the detailed properties
of a system like this one, where short binding regions with asymmetric G-C
content are present, it is important to have a model with
sequence-dependent binding strengths to be able to make more accurate
predictions. Were the G-C pairs in the loop more evenly distributed, we would
expect the results of the average parametrization free energy profile to accurately describe the kissing complex.

\section{Conclusions}

We have extended the nucleotide-level coarse-grained DNA model of Ouldridge
{\it et al.}~\cite{Ouldridge2011} (which distinguishes between A-T and C-G
base-pairing but otherwise treats these interactions at the average base
level) to include sequence-dependent stacking and hydrogen-bonding
interactions. To derive the new parameters, we developed a histogram
reweighting procedure that allowed us to fit to thousands of melting
temperatures of oligomers ranging in length from 6 to 18 base pairs.
Melting temperatures were extracted from SantaLucia's nearest-neighbor
model\cite{SantaLucia1998} which we treat here as a good fit to experiment.

Sequence can have an important effect on melting temperatures. For the same
length oligomer, but different sequences, melting temperatures can differ
by as much as $50\,^{\circ}{\rm C}$. Even for the same sequence content,
but different base-pair ordering, variations in stacking energies mean that
melting temperatures can vary by up to $10\,^{\circ}{\rm C}$. Our new
parametrization reproduces these differences and on average agrees to
within a standard deviation of $0.85\,^{\circ}{\rm C}$ with the SL
nearest-neighbor model. In contrast to the model's ability to capture
thermodynamic properties, our coarse-grained model does not attempt to
include the effects of sequence on structural or mechanical properties of
double-stranded DNA. Instead, these remain as previously reported in
Ref.~\onlinecite{Ouldridge2011}, at least for sequences that are not
extreme in G-C content so that sequence effects average out.

Our new thermodynamic parametrization opens up the possibility of
investigating sequence-dependent DNA phenomena. Specifically, we have
considered here the following five systems:

{\it (a) Heterogeneous stacking transition in single-stranded DNA:} Even
though our stacking parameters do not vary by more than $7 \%$, they can
induce significant spatial and temporal heterogeneity in the stacking of
single strands. For example the difference in stacking probability between
the strongest and the weakest stacking pairs in the oligomer we studied is
large enough that the midpoints of the stacking transition of two separate
pairs in a single strand can be separated by as much as $40\,^{\circ}{\rm
C}$. These results suggest that structural and mechanical properties of
single-stranded DNA should be highly heterogeneous as well.

{\it (b) The hybridization free energy profiles of duplexes:} We studied
three different 12mer sequences at their respective melting temperatures,
finding that sequence heterogeneity also has significant effects on the
probability that the ends of a duplex are open, i.e. that they fray. We
found that A-T ends are typically frayed, while sequences with G-C ends
exhibit a free-energy minimum for a completely closed duplex.

{\it (c) The effect of stacking strength in the loop on hairpin stability:} The SL
model only distinguishes base-pair steps. Given that we used this model to
generate the melting temperatures to which we fit, we were unable to uniquely
isolate the stacking strength of individual base combinations. Additional
experimental data on single-stranded stacking is needed to separate these
interactions. One potential source of data that goes beyond the SL model is
given by experiments on melting of hairpins with poly(dA) and poly(dT)
loops.\cite{Goddard00} By calculating how increasing the stacking strength
in the loop lowers the melting temperatures, we showed that parameters
could be derived that reproduce the expected stronger AA compared to TT
stacking, without significantly changing the quality of our fit to the
overall melting temperatures of duplexes. Nevertheless, we do not yet
include this difference in our parametrization, because to be consistent we
would need similar data to distinguish between other base-pair steps.

{\it (d) The force-extension properties of single strands:} Another
experimental situation where differences in single-stranded stacking have
been measured experimentally is in the force extension of ssDNA. We show
that more strongly stacked sequences should be more extensible for small
forces up to about 5 pN. For certain sequences, experiments have observed a
concave ``plateau'' region in the force-extension curves. We are able to
qualitatively reproduce this feature and, in agreement with previous
explanations,\cite{Seol2007} attribute the plateau region to the different
force response of stiffer stacked and more flexible unstacked regions.
Furthermore, we show that the onset of the plateau region is correlated
with a sharp decrease in the average length of stacked regions with
increasing force. Because the average length of stacked regions drops
rapidly with a relatively small decrease in the average stacking, we argue
that a very large propensity to stack ($> 90$\%) is necessary to give a
similar results to those observed in experiment. We therefore conclude
that if these phenomena are to be explained through largely uncooperative
stacking of bases to form helical ssDNA, as in our model, a high stacking
propensity is required. Furthermore, failure to observe a force plateau
for a sequence does not imply an absence of stacking.

{\it (e) The structure of a kissing-loop complex:} Finally, we applied our
model to study the effect of sequence on the structure of a kissing complex
formed by two hairpins. When the sequences used in the experiments of
Ref.~\onlinecite{Bois2005} are studied, the average base model exhibits one
minimum free-energy structure,\cite{Romano12a} while our sequence-dependent
model also generates a second, qualitatively distinct, stable structure.
The new structure completely exposes a toehold which may significantly
accelerate the DNA catalyst mediated release of free energy stored in the
kissing complex.

The examples described above suggest that our model can be used for many
other DNA applications in nanotechnology and biology where sequence plays a
significant role. Our model should work particularly well for situations
where single-to-double stranded transitions are important. Nevertheless,
users of our model should remain aware of some limitations. Firstly, the
model is only fit to a single salt concentration of $[\mbox{Na}^{+}] =
0.5$M, where the electrostatic properties are strongly screened. A new kind
of parametrization may be necessary to reach significantly lower salt
concentrations. Secondly, the model lacks certain detailed local structural
information, such as major and minor grooving, or sequence dependent
elastic parameters. Furthermore, our model was fit to data that only
includes the effects of base-pair steps. Additional experimental data on
single-stranded stacking is needed to separate out the stacking strength of
individual base combinations. Applications where the effects we neglect are
crucial may therefore be best studied by other models. 

We are developing further improvements to the model, but our work also
highlights the need for new systematic experiments, in particular to
elucidate the basic physics of single-stranded stacking interactions. Such
information would be also of great help to those studying DNA-coated
colloids.~\cite{Mladek2012}

To summarize, we have introduced a new coarse-grained model of DNA that has
been parametrized to reproduce the thermodynamic effects of
sequence-dependent interactions. The current version of the model provides
a computationally efficient and physically accurate tool for the study of
problems ranging from DNA nanotechnology to biology. To facilitate its use,
we have made simulation code implementing Monte Carlo and Brownian dynamics
for the model available as free software called oxDNA at
http://dna.physics.ox.ac.uk.

\section{Acknowledgments} 
The authors would like to thank Erik Winfree, Filip Lanka\v{s}, Felix Ritort and Agnes Noy 
for helpful discussions. The
authors also acknowledge financial support from the Engineering and Physical
Sciences Research Council, University College (Oxford), and from the Oxford Supercomputing Centre for
computer time. P.~\v{S}. is grateful for the award of a Scatcherd
European Scholarship.


\end{document}